\documentclass[11pt]{article}
\usepackage{graphicx}
\usepackage{amsmath,amssymb,amsthm,amsfonts}
\usepackage{color}
\usepackage{amssymb}
\usepackage{cite}
\newtheorem{thm}{Theorem}[section]

\newtheorem{lem}[thm]{Lemma}

\theoremstyle{definition}
\newtheorem{defn}{Definition}[section]
\theoremstyle{remark}
\usepackage{appendix}
\newtheorem{rem}{Remark}[section]

\numberwithin{equation}{section}

\DeclareMathSymbol{\C}{\mathalpha}{AMSb}{"43}

\newcommand{\R}{{\mathbb{R}}}

\def\R{{\mathbb R}}
\def\C{{\mathbb C}}
\allowdisplaybreaks[4]

\newcommand{\bsub}{\begin{subequations}}
\newcommand{\esub}{\end{subequations}$\!$}

\begin{document}

\title{Ground States of Fermionic Nonlinear Schr\"{o}dinger Systems with Coulomb Potential II: The $L^2$-Critical Case}

\author{Bin Chen\thanks{Email:  binchenmath@mails.ccnu.edu.cn.}, \ Yujin Guo\thanks{Email: yguo@ccnu.edu.cn. Y. J. Guo is partially supported by NSFC under Grants 12225106 and 11931012.},\;   and\; Shu Zhang\thanks{Email: szhangmath@mails.ccnu.edu.cn.}\\
\small \it	School of Mathematics and Statistics,\\
\small \it  Key Laboratory of Nonlinear Analysis $\&$ Applications (Ministry of Education),\\
\small \it Central China Normal University, Wuhan 430079, P. R. China\\}


\date{}

\smallbreak \maketitle
\begin{abstract}
As a continuation of  \cite{me}, we consider ground states of the $N$ coupled fermionic nonlinear Schr\"{o}dinger system with a parameter $a $ and the Coulomb potential $V(x)$ in the $L^2$-critical case, where  $a>0$ represents the attractive strength of the  quantum particles. For any given $N\in\mathbb{N}^+$, we prove that  the system admits ground states, if and only if the attractive strength $a$ satisfies $0<a<a^*_N$, where the critical constant $0<a^*_N<\infty$ is the same as the best constant of a dual finite-rank Lieb-Thirring inequality.
By developing the so-called blow-up analysis of many-body fermionic problems, we also prove the mass concentration behavior of ground states for the system as $a\nearrow a_N^*$.
\end{abstract}

\vskip 0.05truein



\vskip 0.2truein

\tableofcontents

\section{Introduction}

It is known (cf. \cite{geomrtric,L3}) that a system of $N$ identical quantum particles  with spin $s$ (such as photons, electrons and neutrons) is usually described by an energy functional of the corresponding $N$-body normalized wave functions  $\Psi\in\bigotimes_{i=1}^NL^2(\R^3, \C^{2s+1})$.  As a continuation of the companion work \cite{me}, in this paper we study ground states of
$N$ spinless (i.e., $s=0$) fermions with the Coulomb potential in the $L^2$-critical case, which can be described by (cf.\cite{geomrtric}) the minimizers of the following constraint variational problem
\begin{equation}\label{1.1}
E_a(N):=\inf\Big\{\mathcal{E}_a(\Psi):\   \|\Psi\|^2_2=1, \ \Psi\in \wedge^NL^2(\R^{3},\C)\cap H^1(\R^{3N},\C)\Big\},\ \ a>0,
\end{equation}
where the energy functional $\mathcal{E}_a(\Psi)$ satisfies
\begin{equation}\label{1.0}
\begin{split}
\mathcal{E}_a(\Psi):=\sum_{i=1}^N\int_{\R^{3N}}\Big(|\nabla_{x_i}\Psi|^2-\sum_{k=1}^K\frac{1}{|x_i-y_k|}|\Psi|^2\Big)dx_1\cdots dx_N -a\int_{\R^{3}}\rho_{\Psi}^{\frac{5}{3}}(x)dx.
\end{split}
\end{equation}
Here $y_1, y_2, \cdots, y_K\in \R^3$ are different from each other, the parameter $a>0$ represents the attractive strength of the  quantum particles, $\wedge^NL^2(\R^3,\C)$ is the subspace of $L^2(\R^{3N},\C)$ consisting of all antisymmetric wave functions, and the one-particle density $\rho_\Psi$ of $\Psi$ is defined as
$$
\rho_\Psi(x):=N\int_{\R^{3(N-1)}}|\Psi(x,x_2\cdots,x_N)|^2dx_2\cdots dx_N.
$$
We refer \cite{ii,i,geomrtric,L3} and the references therein for detailed physical motivations of the variational problem (\ref{1.1}).

Following the spectral theorem (see \cite{i} and the references therein), we denote the non-negative self-adjoint operator $\gamma=\sum_{i=1} ^Nn_i|u_i\rangle \langle u_i|$ on $L^2(\R^3,\C)$ by
\begin{equation}\label{gamma}
\gamma\varphi(x)=\sum_{i=1}^Nn_iu_i(x)(\varphi, u_i)_{L^2(\R^3, \C)},\ \ \forall\ \varphi\in L^2(\R^3, \C),
\end{equation}
where both $n_i\geq0$ and $u_i\in L^2(\R^3, \C)$ hold for $i=1,\, \cdots, N$. Moreover, we use
\begin{equation}\label{rho}
\rho_\gamma(x)=\sum_{i=1}^Nn_i|u_i(x)|^2
\end{equation}
to denote the corresponding density of $\gamma$. The similar argument of \cite[Appendix A and Lemma 2.3]{me} then yields that the problem  \eqref{1.1} can be reduced equivalently to the following form
\begin{equation}\label{1.2}
\begin{split}
E_a(N)=&\inf\Big\{\mathcal{E}_a(\gamma):\, \gamma=\sum_{i=1} ^N|u_i\rangle \langle u_i|,\  u_i\in H^1(\R^3,\R),\Big.\\
&\Big. \ \ \ \  \ \ \  (u_i,u_j)_{L^2}=\delta_{ij},\  i,j=1,\cdots,N\Big\},\, \ a>0,\, \ N\in \mathbb{N}^+,
\end{split}
\end{equation}
where the energy functional $\mathcal{E}_a(\gamma)$ satisfies
\begin{equation}\label{1.3a}
\mathcal{E}_a(\gamma):=\mathrm{Tr}\big(-\Delta+V(x)\big) \gamma-a\int_{\R^3}\rho_\gamma^{\frac{5}{3}}dx,
\end{equation}
and the function $V(x)\le 0$ is the Coulomb potential of the form
\begin{equation}\label{v}
V(x):=-\sum_{k=1}^K\frac{1}{|x-y_k|}
\end{equation}
containing different singular points $y_1,\cdots ,  y_K\in \R^3$. Throughout the present whole paper we therefore focus on the analysis of \eqref{1.2}, instead of \eqref{1.1}. As a continuation of  \cite{me}, which handles the $L^2-$subcritical case of $E_a(N)$, the main purpose of the present paper is to analyze the limiting concentration behavior of minimizers for the variational problem $E_a(N)$ defined in \eqref{1.2}.

For any given $N\in\mathbb{N}^+$, we now  consider the following minimization problem
\begin{equation}\label{k}
\begin{split}
0<a^*_N:=&\inf\Big\{\frac{\|\gamma\|^{\frac{2}{3}}\mathrm{Tr}(-\Delta \gamma)}{\int_{\R^3}\rho_\gamma^{5/3}dx}:\ \gamma=\sum_{i=1} ^Nn_i|u_i\rangle \langle u_i|\neq0, \ u_i\in H^1(\R^3), \ n_i\geq0\Big\},
\end{split}
\end{equation}
where $\rho_\gamma$ is as in (\ref{rho}), and $\|\gamma\|>0$ denotes the norm of the operator $\gamma$.
The proof of   \cite[Theorem 6]{ii} gives essentially that for any $N\in\mathbb{N}^+$, the best constant $a^*_N\in(0,+\infty)$ of (\ref{k}) is attained, and any minimizer $\gamma^{(N)}$ of $a^*_N$ can be written in the form
\begin{equation}\label{form}
	\gamma^{(N)}=\|\gamma^{(N)}\|\sum_{i=1}^{R_N}|Q_i\rangle\langle Q_i|, \ \   (Q_i,Q_j)=\delta_{ij}\ \ \mathrm {for}\ \ i,j=1,\cdots, R_N,
\end{equation}
where the positive integer $R_N\in[1, N]$, and
the orthonormal  family $Q_1, \cdots,Q_{R_N}$ solves the following fermionic nonlinear Schr\"{o}dinger system
\begin{equation}\label{nls}
\begin{split}
\Big[-\Delta-\frac{5}{3}a^*_N\Big(\sum_{j=1}^{R_N}Q_j^2\Big)^{\frac{2}{3}} \Big]Q_i	&=\hat{\mu}_i Q_i\ \ \mathrm{in}\, \ \R^3, \ \ i=1,\cdots, R_N.
\end{split}
\end{equation}
Here $\hat{\mu}_1<\hat{\mu}_2\leq\cdots \leq\hat{\mu}_{R_N}<0$ are the $R_N$ first eigenvalues (counted with multiplicity) of the operator
$$\hat H_{\gamma^{(N)}}:=-\Delta-\frac{5}{3}a^*_N\Big(\sum_{j=1}^{R_N}Q_j^2\Big)^{\frac{2}{3}}\ \ \ \mbox{in} \ \,\R^3.$$
Particularly, the authors proved in \cite[Proposition 11]{ii} that
\begin{equation}\label{2A:strict}
\mbox{$a_N^*>a_{2N}^*$ holds for any $N\in\mathbb{N}^+$,}
\end{equation}
and thus there exists an infinite sequence of integers
$N_1=1<N_2=2<N_3<\cdots$
such that
\begin{equation}\label{strict}
a^*_{N_m-1}>a^*_{N_m},\ \ \ m=2,3,4,\cdots,
\end{equation}
which further  implies that any minimizer $\gamma^{(N_m)}$ of $a^*_{N_m}$ satisfying (\ref{strict}) must satisfy $\mathrm{Rank}\big(\gamma^{(N_m)}\big)=N_m$.
We also comment that the uniqueness of minimizers for $a^*_{N}$ is still open for any $N\geq 2$.

The minimization problem $E_a(1)$ defined in \eqref{1.2} is essentially an $L^2$-critical constraint variational problem, which was investigated widely over the past few years, starting from the earlier work \cite{cr3}. Moreover, we expect that  for any $N\in\mathbb{N}^+$, the minimizers of $E_a(N)$ are connected with ground states of a  fermionic nonlinear Schr\"{o}dinger system, in the sense that

\begin{defn}\label{dfn:1} (\emph{Ground states}). A system $(u_1,\cdots, u_N )\in \big(H^1(\R^{3},\R )\big)^N$ with $(u_i, u_j)_{L^2}$ $=\delta_{ij}$ is called a ground state of
\begin{equation}\label{1.3M}
H_V u_i:=\Big[-\Delta +V(x)- \frac{5a}{3}\Big(\sum_{j=1}^Nu_j^2\Big)^{\frac{2}{3}}\Big]u_i=\mu_i u_i\ \ \mathrm{in}\ \ \R^3,\, \ i=1,\cdots,N,
\end{equation}
if it solves the system \eqref{1.3M}, where $\mu_1<\mu_2\leq\cdots\leq\mu_N\leq0$ are the $N$ first eigenvalues (counted with multiplicity) of the operator $H_V$.
\end{defn}

\subsection{Main results}

The purpose of this subsection is to introduce the main results of the present paper.
Motivated by \cite{ii, cr3}, in this paper we first prove the following existence and  nonexistence of minimizers for $E_a(N)$.

\begin{thm}\label{th1} For any fixed $N\in\mathbb{N}^+$, suppose $E_a(N)$ is defined in \eqref{1.2}, and let $0<a^*_N<\infty$ be defined by \eqref{k}. Then
\begin{enumerate}
\item  If $0<a<a^*_N$, then $E_a(N)$ admits at least one minimizer. Moreover, any minimizer $\gamma$ of $E_a(N)$ satisfies $\gamma=\sum_{i=1} ^N|u_i\rangle \langle u_i|$, where $(u_1, \cdots, u_N)$ is a ground state of the following fermionic nonlinear Schr\"{o}dinger system
\begin{equation}\label{1.3}
H_V u_i:=\Big[-\Delta +V(x)- \frac{5a}{3}\Big(\sum_{j=1}^Nu_j^2\Big)^{\frac{2}{3}}\Big]u_i=\mu_i u_i\ \ \text{in}\ \ \R^3,\, \ i=1,\cdots,N.
\end{equation}
Here the Coulomb potential $V(x)\le 0$ is as in (\ref{v}), and $\mu_1 < \mu_2\leq\cdots\leq\mu_N< 0$ are the $N$ first eigenvalues, counted with multiplicity, of the operator $H_V$ in $\R^3$.
\item  If $a\geq a^*_N$, then $E_a(N)$ does not admit any minimizer, and $E_a(N)=-\infty$.
\end{enumerate}
\end{thm}

We remark that for any fixed $N\in\mathbb{N}^+$, Theorem \ref{th1} provides a complete classification on the existence and nonexistence of minimizers for $E_a(N)$ in terms of $a>0$. Moreover, the proof of Theorem \ref{th1} implies that Theorem \ref{th1} can be naturally extended not only to  the general singular potential $V(x)=-\sum_{k=1}^K|x-y_k|^{-s_k}$ with $0<s_k<2$, but also to  the generally dimensional case $\R^d$ with $d\geq3$, if the exponent $\frac{5}{3}$ in the last term of \eqref{1.3a} is replaced by $1+\frac{2}{d}$. On the other hand, in order to prove Theorem \ref{th1}, in Section 2 we shall derive the boundedness, monotonicity and some other analytical properties of the energy $E_a(N)$ in terms of $N>0$ and $a>0$.
Furthermore, the existence proof of Theorem \ref{th1}  is based on an adaptation of the classical concentration compactness principle (cf. \cite{concern}\cite[Section 3.3]{begain}). Towards this purpose, the  key step  is to prove in Subsection 2.1
that  the following strict inequality holds for any fixed $0<\lambda< N$,
\begin{equation}\label{1.3MM}
E_a(N)<E_a(\lambda), \ \ \mbox{if \ $E_a(\lambda)$\, admits\, minimizers,}
\end{equation}
where $E_a(\lambda)$ is  defined by \eqref{2.0} below, see  (\ref{2.57A}) for more details.


By developing the so-called blow-up analysis of many-body fermionic problems, in the following we focus on exploring the limiting concentration behavior of  minimizers for $E_a(N)$
as $a \nearrow a^*_N$. For simplicity, we first address the particular case $N=2$, where $a^*_1>a^*_2$ holds true in view of (\ref{2A:strict}).

\begin{thm}\label{th3}
Let $\gamma_a=\sum_{i=1}^2|u_i^a\rangle\langle u_i^a|$ be a minimizer of $E_a(2)$ for $0<a<a^*_2$. Then for any sequence $\{a_n\}$ satisfying $a_n\nearrow a^*_2$ as $n\rightarrow\infty$, there exist a subsequence, still denoted by $\{a_n\}$,  of $\{a_n\}$ and a point  $y_{k_*}\in\{y_1,\cdots,y_K\}$ given by \eqref{v} such that
\begin{equation}\label{th1.3.1}
\begin{split}
w_i^{a_n}(x):&=\epsilon_{a_n}^{\frac{3}{2}}u_i^{a_n}\big(\epsilon_{a_n}x+y_{k_*}\big)\\[1.5mm]
&\to w_i(x)\ \ \text{strongly\ in}\ \ H^1(\R^3)\cap L^\infty(\R^3)\ \ \text{as}\ \ n\rightarrow\infty, \ \ i=1,2,
\end{split}
\end{equation}
and
\begin{equation}\label{th1.3.2}
\lim\limits_{n\to\infty}\epsilon_{a_n}E_{a_n}(2)=-\int_{\R^3}\rho^{5/3}_{\gamma}dx,
\end{equation}
where $\epsilon_{a_n}:=a^*_2-a_n>0$, and $\gamma:=\sum_{i=1}^2|w_i\rangle\langle w_i|$ satisfying $(w_i,w_j)=\delta_{ij}$ is an optimizer of $a^*_2$.	
\end{thm}


\begin{rem}\label{rem1}
$(1)$. As  a byproduct of Theorem \ref{th3}, we shall derive in \eqref{3.38} that the following  identities hold true:
\begin{equation}
\frac{1}{a_2^*}\mathrm{Tr}\big(-\Delta \gamma\big)=\int_{\R^3}\rho^{5/3}_{\gamma}dx=\frac{1}{2}\int_{\R^3}|x|^{-1}\rho_{\gamma}dx,
\end{equation}
where $\gamma=\sum_{i=1}^2|w_i\rangle\langle w_i|$ given by \eqref{th1.3.1} is an optimizer of $a^*_2$.

$(2)$. The proof of Theorem \ref{th3} shows that Theorem \ref{th3} actually holds true for any $E_a(N)$, provided that $2\leq N\in\mathbb{N}^+$ satisfies $a^*_{N-1}>a^*_{N}$, see Remark \ref{rem3.1} for more details.

$(3)$.  The $L^\infty$-uniform convergence (\ref{th1.3.1}) presents the following mass concentration behavior of minimizers $\gamma_{a_n}=\sum_{i=1}^2|u_i^{a_n}\rangle\langle u_i^{a_n}|$ for  $E_{a_n}(2)$  as $a_n\nearrow a^*_2$:
\begin{equation}\label{1.19}
\gamma_{a_n}(x,y)\approx \big(a^*_2-a_n\big)^{-3}\gamma\Big(\frac{x-y_{k_*}}{a^*_2-a_n}, \frac{y-y_{k_*}}{a^*_2-a_n}\Big) \ \ \mbox{as}\ \ a_n\nearrow a^*_2,
\end{equation}
where $\gamma(x,y)=\sum_{i=1}^2w_i(x)w_i(y)$ denotes the integral kernel of $\gamma$, and $y_{k_*}\in\{y_1,\cdots, y_K\}$ is as in Theorem \ref{th3}.
This implies that the mass of the minimizers for $E_{a_n}(2)$  as $a_n\nearrow a^*_2$ concentrates at a global minimum point $y_{k_*}\in\{y_1,\cdots, y_K\}$ of the Coulomb potential $V(x)=-\sum_{k=1}^K|x-y_k|^{-1}$. It is still interesting to further address the exact point $y_{k_*}$ among the set $\{y_1,\cdots, y_K\}$.
\end{rem}

We now follow three steps to explain briefly the general strategy of proving Theorem \ref{th3}:

The first step of proving Theorem \ref{th3} is to derive the precise upper bound (\ref{d3}) of the energy $E_a(2)$ as $a\nearrow a^*_2$, which further implies the estimates of Lemma \ref{lem3.1}. Due to the orthonormal constrained conditions, it however seems difficult to borrow the existing methods (e.g. \cite{cr3}) of analyzing the $L^2-$critical variational problems. In order to overcome this difficulty, in Lemma \ref{lem3.1} we shall construct a new type of test operators involved with the complicated analysis.

As the second step of proving Theorem \ref{th3}, we shall prove Lemma \ref{lem3.2} on the $H^1$-uniform convergence of  the  sequence $\{w_i^{a_n}\}_n$ as $a_n\nearrow a^*_2$ for $ i=1, 2$, where $w_i^{a_n}$ is defined by
\begin{equation}\label{0:ww}
w_i^{a_n}(x):=\epsilon_{a_n}^{\frac{3}{2}}u_i^{a_n}\big(\epsilon_{a_n}x+y_{k_*}\big),\ \ \epsilon_{a_n}:=a^*_2-a_n>0,
\end{equation}
and $\gamma_{a_n}=\sum_{i=1}^2|u_i^{a_n}\rangle\langle u_i^{a_n}|$ is a minimizer of $E_{a_n}(2)$.
To reach this aim, in Section 3 we shall apply the following finite-rank Lieb-Thirring inequality
\begin{equation}\label{lt}
	L^*_{N}\int_{\R^3}W^{\frac{5}{2}}(x)dx\geq \sum_{i=1}^N\big|\lambda_i(-\Delta-W(x))\big|, \ \ \ \forall\ 0\leq W(x)\in L^{\frac{5}{2}}(\R^3)\backslash\{0\},
\end{equation}
where the best constant $L^*_{N}\in(0,+\infty)$ is attainable (cf.\cite[Corollary 2]{iii}), and $\lambda_i\big(-\Delta-W(x)\big)\leq0$ denotes the $i$th negative eigenvalue (counted with multiplicity) of $-\Delta-W(x)$ in $L^2(\R^3)$ when it exists, and zero otherwise. We point out that by proving
\begin{equation}\label{dual}	 a^*_N\big(L^*_{N}\big)^{\frac{2}{3}}=\frac{3}{5}\Big(\frac{2}{5}\Big)^{\frac{2}{3}},
\end{equation}
it was addressed in \cite{ii, iii,open}  that  the corresponding inequality of \eqref{k} is dual to the finite rank Lieb-Thirring inequality \eqref{lt}. Applying the energy estimates of the first step, together with the above dual relationship and  the strict inequality $a^*_{1}>a^*_{2}$, we shall prove the crucial  $L^{\frac{5}{3}}$-uniform convergence of the density sequence $\{\rho_n\}:=\big\{\sum_{i=1}^2|w_i^{a_n}|^2\big\}$ as $a_n\nearrow a^*_2$. Since any minimizer $\gamma^{(2)}$ of $a^*_{2}$  satisfies $\gamma^{(2)}=\|\gamma^{(2)}\|\sum_{i=1}^2|Q_i\rangle\langle Q_i|$ with $(Q_i,Q_j)=\delta_{ij}$, in Lemma \ref{lem3.2} we are then able to establish finally the $H^1$-uniform convergence of $\{w_i^{a_n}\}_n$ as $a_n\nearrow a^*_2$.

The third step of proving Theorem \ref{th3} is to establish the energy estimate \eqref{th1.3.2} and the $L^\infty$-uniform convergence of $\{w_i^{a_n}\}_n$ as $a_n\nearrow a^*_2$. Actually, employing the energy estimates  and the  $L^{\frac{5}{3}}$-uniform convergence of the previous two steps,  we can analyze the exact leading term of $E_{a_n}(2)$ as $a_n\nearrow a_2^*$, which then helps us prove the energy limit \eqref{th1.3.2}. On the other hand, to prove the $L^\infty$-uniform convergence of $\{w_i^{a_n}\}_n$, we shall prove the following  uniformly exponential decay
\begin{equation}\label{th1.3.3}
\Big(|w_1^{a_n}(x)|^2+|w_2^{a_n}(x)|^2\Big)\leq C(\theta)e^{-\theta|x|}\ \ \text{uniformly\ in}\ \, \R^3\ \ \text{as}\ \ n\rightarrow\infty,
\end{equation}
where the constants $\theta>0$ and $C(\theta)>0$ are independent of $n>0$, see Lemma \ref{3.3} for more details. Unfortunately, the exponential decay \eqref{th1.3.3} cannot be  established by the standard comparison principle, due to the singularity of the Coulomb potential $V(x)$. For this reason,  we shall prove \eqref{th1.3.3} by employing Green's  functions. The complete  proof of  Theorem \ref{th3} is given in Section 3.

In order to discuss the general case of $E_a(N)$, we next analyze the limiting concentration behavior of minimizers for $E_a(3)$ as $a \nearrow a^*_3$, where $a^*_1>a^*_2\geq a^*_3$ holds true in view of \eqref{k} and (\ref{2A:strict}).


\begin{thm}\label{th4}
Let $\gamma_a=\sum_{i=1}^3|u_i^a\rangle\langle u_i^a|$ be a minimizer of $E_a(3)$ for $0<a<a^*_3$. Then for any sequence $\{a_n\}$ satisfying $a_n\nearrow a^*_3$ as $n\rightarrow\infty$, there exist a subsequence, still denoted by $\{a_n\}$,  of $\{a_n\}$ and a point  $y_{k_*}\in\{y_1,\cdots,y_K\}$ given by \eqref{v} such that
\begin{equation}\label{th4.1}
w_i^{a_n}(x):=\epsilon_{a_n}^{\frac{3}{2}}u_i^{a_n}\big(\epsilon_{a_n}x+y_{k_*}\big)\to w_i\ \ \text{strongly\ in}\ \ L^\infty(\R^3)\ \ \mathrm{as}\ \ n\rightarrow\infty,\ \ i=1, 2, 3,
\end{equation}
where $\epsilon_{a_n}:=a^*_3-a_n>0$, and $\gamma:=\sum_{i=1}^3|w_i\rangle\langle w_i|$ is an optimizer of $a^*_3$. 
 Furthermore, we have
\begin{enumerate}
\item  If either $a_2^*>a^*_3$, or $a_2^*=a^*_3$ and $Rank(\!\gamma) =3$, then $(w_i, w_j)=\delta_{ij}$ holds for $i, j=1,2,3$.
\item  If $a_2^*=a^*_3$ and  $Rank(\!\gamma) =2$, then  \\
Case 1: Either $w_i(x)\not\equiv0$ in $\R^3$ holds for all $i=1,2,3$, where $\|w_1\|_2^2=1$, $(w_1,w_2)=(w_1,w_3)=0$, and
\begin{equation}\label{th4.1:M}
w_3(x)\equiv\pm\sqrt{\|w_2\|_2^{-2}-1}\ w_2(x) \ \ \text{in}\ \  \R^3.
\end{equation}
Case 2: Or  $(w_i, w_j)=\delta_{ij}$ holds for $i, j\in \{1,2\}$, and $w_{3}(x)\equiv0$ in $\R^3$.
\end{enumerate}
\end{thm}

\begin{rem}\label{rem2}
$(1).$ As for Theorem \ref{th4} (1), the similar proof of Theorem \ref{th3} can further yield the $H^1$-strong convergence of $\{w_i^{a_n}\}$ defined in (\ref{th4.1}) as $n\to\infty$, where $i=1, 2, 3$.

$(2).$ We expect that the argument of Theorem \ref{th4}  can further yield the following general results: For any given $N\ge 3$, if $\gamma_{a_n}=\sum_{i=1}^N|u_i^{a_n}\rangle\langle u_i^{a_n}|$ is a minimizer of $E_{a_n}(N)$, where $a_n\nearrow a^*_N$ as $n\rightarrow\infty$, then there exist  a point  $y_{k_*}\in\{y_1,\cdots,y_K\}$ and a minimizer $\gamma:=\sum_{i=1}^N|w_i\rangle\langle w_i|$  of $a^*_N$ such that, up to a subsequence if necessary,
\begin{equation}\label{1:rem1.2}
w_i^{a_n}(x):=\epsilon_{a_n}^{\frac{3}{2}}u_i^{a_n}\big(\epsilon_{a_n}x+y_{k_*}\big)\to w_i\ \ \text{strongly\ in}\ \, L^\infty(\R^3)\ \ \mathrm{as}\ \ n\rightarrow\infty
\end{equation}
holds for $i=1, 2,\cdots,N$, where Rank$(\gamma)=\text{dim}\big(\text{span}\{w_1,\cdots,w_N\}\big)$ $:=R_N\in [2, N]$. Moreover,  either $w_i\not\equiv 0$ in $\R^3$ holds for all $i=1, 2, \cdots, N$, or there exists exactly one integer $M\in [1, N-2]$  such that $w_{i_1}(x)\equiv\cdots\equiv w_{i_M}(x)\equiv0$ in $\R^3$, in which case it necessarily has
$$\big\{i_1, i_2, \cdots, i_M\big\}=\big\{N-M+1, N-M+2,\cdots, N\big\}.$$
\end{rem}

We next explain, totally by four steps, the general strategy of proving Theorem \ref{th4}, which can be summarized as
{\em the blow-up analysis of many-body fermionic problems}. The first three steps of proving Theorem \ref{th4} are similar to those of proving Theorem \ref{th3}, which then yield  the $L^\infty$-uniform convergence of \eqref{th4.1} in view of \eqref{2A:strict}.
Moreover, since Theorem \ref{th4} (1) focuses essentially on the cases where $\text{dim}\, \text{span}\{w_1^{a_n},w_2^{a_n},w_3^{a_n}\}=\text{dim}\, \text{span}\{w_1,w_2,w_3\}$, and $\sum_{i=1}^3|w_i\rangle\langle w_i|$ is an optimizer of $a^*_3$, where $w_i$ is as in \eqref{th4.1}, the above analysis procedure also yields that $(w_i, w_j)=\delta_{ij}$,  $i, j=1,2,3$, holds in this case. This proves Theorem \ref{th4} (1).

The fourth step of proving Theorem \ref{th4} is to complete the proof of  Theorem \ref{th4} (2), which is devoted to the situation where $a_2^*=a^*_3$ and $Rank(\!\gamma) =2$. Under the assumptions of Theorem \ref{th4} (2), we first note that
\begin{equation}\label{ess}
 \text{dim}\, \text{span}\big\{w_1^{a_n},w_2^{a_n},w_3^{a_n}\big\}>\text{dim}\, \text{span}\{w_1,w_2,w_3\},
\end{equation}
where $w_i$ is given by \eqref{th4.1}, and $\gamma:=\sum_{i=1}^3|w_i\rangle\langle w_i|$ is an optimizer of $a^*_3$.
This implies that the components of  the system $(w_1,w_2,w_3)$ are linearly dependent. Since $w_1^{a_n}(x)=\epsilon_{a_n}^{\frac{3}{2}}u_1^{a_n}\big(\epsilon_{a_n}x+y_{k_*}\big)$ and $u_1^{a_n}$ is the first eigenfunction of the operator $H_V^{a_n}$ given by \eqref{1.3}, we shall prove that
\begin{equation}\label{1:aadd}
\mbox{$\| w_1\|_2=1$ and $(w_{1},w_i)=0$ holds for $i=2,3$.}
\end{equation}

Following \eqref{form}, Case 1 of Theorem \ref{th4} (2) can be directly proved in view of above facts. As for Case 2 of Theorem \ref{th4} (2), the challenging point is to further show that $w_{2}(x)\not\equiv0$ in $\R^3$. We shall prove this result as follows. Similar to the proof of Theorem \ref{th3}, one can establish the following convergence:
\begin{eqnarray}\label{1.25}
	|\nabla w_i^{a_n}|\rightarrow |\nabla w_i|\ \ \ \text{strongly\  in} \, \ L^2(\R^3)\ \ \text{as} \ \ n\to\infty,\ \    i=1,2,3,
\end{eqnarray}
and
\begin{eqnarray}\label{1.26}
	\sum_{i=1}^3|w_i^{a_n}|^2\rightarrow \sum_{i=1}^3w_i^2\ \ \ \text{strongly\  in}\  L^{\frac{5}{3}}(\R^3)\ \text{ as}\,\ n\to\infty.
\end{eqnarray}
Note from Theorem \ref{th1} that the minimizer $\gamma_{a_n}=\sum_{i=1}^3|u_i^{a_n}\rangle\langle u_i^{a_n}|$ satisfies the following system
\begin{equation}\label{1.26M}
	 H_V^{a_n}u_i^{a_n}:=\Big[-\Delta+V(x)-\frac{5}{3}a_n\Big(\sum_{j=1}^3|u_j^{a_n}|^2\Big)^{\frac{2}{3}}\Big]u_i^{a_n}
	=\mu_i^{a_n}u_i^{a_n}\ \ \  \mathrm{in}\ \ \R^3,\ \   i=1,2,3,
\end{equation}
where $\mu_1^{a_n}<\mu_2^{a_n}\leq\mu_3^{a_n}<0$   are the $3$-first eigenvalues (counted with multiplicity) of the operator $H_V^{a_n}$. By contradiction, we next suppose that $w_{2}(x)\equiv0$ in $\R^3$.
By deriving some energy estimates of $u_i^{a_n}$ and $\mu_i^{a_n}$ as $n\to\infty$, we shall derive from \eqref{th4.1}, (\ref{1.25})--\eqref{1.26M} that
\begin{equation}\label{1.26a}
	\lim\limits_{n\to\infty}\sum_{i=1}^3\epsilon_{a_n}^2\mu_i^{a_n}=
	\lim\limits_{n\to\infty}\sum_{i\neq2}^3\epsilon_{a_n}^2\mu_i^{a_n}, \ \ \lim\limits_{n\to\infty}\epsilon_{a_n}^2\mu_2^{a_n}<0,
\end{equation}
a contradiction, which yields that $w_{2}(x)\not\equiv0$ in $\R^3$. Applying (\ref{1:aadd}), this completes the proof of Theorem \ref{th4} (2).  For the detailed proof of Theorem \ref{th4}, we refer the reader to Section 4.

This paper is organized as follows. Section 2 is devoted to the proof of Theorem \ref{th1} on the existence and nonexistence of minimizers for $E_a(N)$. In Section 3, we shall  address Theorem \ref{th3} on the limiting concentration behavior  of  minimizers for $E_a(2)$ as $a \nearrow a^*_2$. The proof of Theorem \ref{th4} is given in Section 4, which is concerned with the limiting concentration behavior  of  minimizers for $E_a(N)$ in the case $N=3$.

\section{Existence of Minimizers for $E_a(N)$}

In this section, we mainly prove Theorem \ref{th1} on the existence and nonexistence of minimizers for $E_a(N)$ defined by \eqref{1.2}, where $N\in\mathbb{N}^+$ is arbitrary. Towards this purpose, we need to introduce the following general minimization problem
\begin{align}\label{2.0}
E_a(\lambda):=&\inf\Big\{  \mathcal{E}_a(\gamma):\ \gamma=\sum_{i=1}^{N'}|u_i\rangle\langle u_i|+(\lambda-N')|u_{N'}\rangle\langle u_{N'}|,\Big.\nonumber\\
&\ \ \ \ \ \ \ \ u_i\in H^1(\R^3),\  (u_i,u_j)=\delta_{ij},\ i,j=1,\cdots,N'\Big\},\,\ a>0,\,\ \lambda>0,
\end{align}
where the energy functional $\mathcal{E}_a(\gamma)$ is defined by \eqref{1.3a}, and $N'$ is the smallest integer such that $\lambda\leq N'$.
One can note that when $\lambda=N'\in\mathbb{N}^+$, \eqref{2.0} coincides with $E_a(N')$ defined in  \eqref{1.2}.

We first address the analytical properties of $E_a(\lambda)$. Denoting $\mathcal{B}\big(L^2(\R^3)\big)$  the set of bounded linear operators on $L^2(\R^3)$, we then have the following equivalence of $E_a(\lambda)$.

\begin{lem}\label{lem:2.0}  Suppose the problem $E_a(\lambda)$ is defined by (\ref{2.0}) for $a>0$ and $\lambda>0$. Then we have
\begin{equation}\label{equ}
E_a(\lambda)=\inf\limits_{\gamma'\in\mathcal{K}_\lambda}\mathcal{E}_a(\gamma'),
\end{equation}
where the functional $\mathcal{E}_\alpha(\gamma')$ is as in (\ref{1.3a}), and $\mathcal{K}_{\lambda}$ is defined by
\begin{equation}\label{set}
\mathcal{K}_{\lambda}=\big\{\gamma'\in \mathcal{B}\big(L^2(\R^3)\big):\ 0\leq\gamma'=(\gamma')^*\leq 1,\ \mathrm{Tr}\gamma'=\lambda,\ \mathrm{Tr}(-\Delta \gamma')<\infty\big\}.
\end{equation}
Moreover,  if $\inf\limits_{\gamma'\in\mathcal{K}_\lambda}\mathcal{E}_a(\gamma')$ admits minimizers, then $E_a(\lambda)$ also admits minimizers.
\end{lem}

Since Lemma \ref{lem:2.0} can be proved in a similar approach of \cite[Lemma 11]{i} and \cite[Lemma 2.3]{me}, we omit the detailed proof for simplicity. Applying the equivalent version (\ref{equ}) of $E_a(\lambda)$, one can obtain the following properties of  $E_a(\lambda)$.

\begin{lem}\label{lem2.1}
For any fixed $N\in\mathbb{N}^+$, suppose the  constant $0<a^*_N<\infty$ is defined by \eqref{k}. Then the energy $E_a(\lambda)$ defined in \eqref{2.0} admits the following properties:
\begin{enumerate}
\item If $0< a<a^*_N$, then $-\infty<E_a(\lambda)<0$ holds for all $\lambda\in(0,N]$.

\item   If $0< a\leq a^*_N$, then  $E_a(\lambda)$ is decreasing in $\lambda\in (0,N]$.

\item If $a\geq a^*_N$, then $E_a(N)=-\infty$.
\end{enumerate}
\end{lem}

\noindent \textbf{Proof.} 1. For $\lambda\in(0,N]$,
set
\begin{equation}\label{ga}
\gamma:=\sum_{i=1}^{N'}|u_i\rangle\langle u_i|+(\lambda-N')|u_{N'}\rangle\langle u_{N'}|,\ \ u_i\in H^1(\R^3),\ \ (u_i, u_j)_{L^2}=\delta_{ij},
\end{equation}
where $N'$ is the smallest integer such that $\lambda\leq N'$. Since $\lambda\in(0,N]$, we have $N'\leq N$. By the definition of $a^*_N$ defined in (\ref{k}), we then get that
\begin{equation}\label{2.1b}
	\mathrm{Tr}(-\Delta \gamma)\geq\|\gamma\|^{\frac{2}{3}}\mathrm{Tr}(-\Delta \gamma)\geq a^*_N\int_{\R^3}\rho_\gamma^{\frac{5}{3}}dx,
\end{equation}
where $\rho_\gamma$, representing the density of $\gamma$, is defined by \eqref{rho}.
By Hardy's inequality, we have
\begin{equation}\label{2:H}
|x|^{-1}\leq \varepsilon \big(-\Delta\big)+4\varepsilon^{-1},\,  \ \mbox{where}\,\ \varepsilon>0 \,  \ \mbox{is\ arbitrary}.
\end{equation}
It then yields that
\begin{equation}\label{ep}
V(x)=-\sum_{k=1}^K|x-y_k|^{-1}\geq-\varepsilon K(-\Delta)-4\varepsilon^{-1}K \ \ \mathrm{holds \ for\  any}\  \varepsilon>0.
\end{equation}
For simplicity, we denote $n_1=\cdots=n_{N'-1}=1$ and  $n_{N'}=\lambda-N'+1$, so that \begin{equation}\label{2.7A}
\gamma=\sum_{i=1}^{N'}n_i|u_i\rangle\langle u_i|,\ \ \rho_\gamma=\sum_{i=1}^{N'}n_i|u_i|^2.
\end{equation}
By the definitions of \eqref{gamma} and  Trace, it  then follows from  \eqref{ep} that
\begin{equation}\label{trace}
\begin{split}
\mathrm{Tr}\big(-\Delta+V(x)\big)\gamma
&=\sum_{i=1}^{N'}n_i\Big(\big(-\Delta+V(x)\big)u_i, u_i\Big)\\
&\geq(1-\varepsilon K)\sum_{i=1}^{N'}n_i\big(-\Delta u_i, u_i\big)-4\varepsilon^{-1}K\sum_{i=1}^{N'}n_i( u_i, u_i)\\
&=(1-\varepsilon K)\mathrm{Tr}(-\Delta \gamma)-4\varepsilon^{-1}K\lambda,\ \ \  \varepsilon>0.
\end{split}
\end{equation}
For $0\leq a<a^*_N$, taking $\varepsilon>0$ so that $\varepsilon K=\frac{1}{2}\Big(1-\frac{a}{a^*_N}\Big)>0$,  we further obtain from \eqref{2.1b} and  \eqref{trace} that
\begin{equation}\label{2.2}
\begin{split}
\mathcal{E}_a(\gamma)=&\mathrm{Tr}\big(-\Delta+V(x)\big) \gamma-a\int_{\R^3}\rho_\gamma^{\frac{5}{3}}dx\\
\geq&\ (1-\varepsilon K)\mathrm{Tr}(-\Delta \gamma)-\frac{a}{a^*_N}\mathrm{Tr}(-\Delta \gamma)-4\varepsilon^{-1}K\lambda\\
=&\ \frac{1}{2}\Big(1-\frac{a}{a^*_N}\Big)\mathrm{Tr}(-\Delta \gamma)
-\frac{8\lambda K^2a^*_N}{a^*_N-a}\geq-\frac{8\lambda K^2a^*_N}{a^*_N-a}.
\end{split}
\end{equation}
Since $\gamma$ is arbitrary, we obtain from \eqref{2.2} that $E_a(\lambda)>-\infty$ holds for any $0<\lambda\leq N$ and $0\leq a<a^*_N$.

Associated to (\ref{ga}), we define
\begin{equation}\label{2.3M}
\gamma_t:=\sum_{i=1}^{N'}t^3|u_i(t\cdot)\rangle\langle u_i(t\cdot)|+(\lambda-N')t^3|u_{N'}(t\cdot)\rangle\langle u_{N'}(t\cdot)|,\ \ t>0.
\end{equation}
Similar to the first identity of \eqref{trace}, one can calculate from (\ref{v}) that
\begin{equation}\label{2.3}
\begin{split}
\mathcal{E}_a(\gamma_t)
&= t^2\mathrm{Tr}(-\Delta\gamma)
-at^2\int_{\R^3}\rho_{\gamma}^{\frac{5}{3}}dx
-t\sum_{k=1}^K\int_{\R^3}|x-ty_k|^{-1}\rho_{\gamma}dx,
\end{split}
\end{equation}
where $\gamma$ and $\rho_\gamma$ are as in (\ref{2.7A}).
Since
\begin{eqnarray*}
\lim\limits_{t\rightarrow0}\int_{\R^3}|x-ty_k|^{-1}\rho_{\gamma}dx
=\int_{\R^3}|x|^{-1}\rho_{\gamma}dx>0,\ \ k=1,\cdots, K,
\end{eqnarray*}
we obtain from \eqref{2.3} that if $0\leq a<a^*_N$ and $0<\lambda \leq N$, then $\mathcal{E}_a(\gamma_t)<0$ holds for sufficiently small $t>0$. This further implies that $E_a(\lambda)<0$ holds for any $0\leq a<a^*_N$ and $0<\lambda \leq N$.

2. For any given $0<\lambda_1<\lambda\leq N$,  consider any operators $\gamma_1$ and $\gamma_2$ satisfying the following constraint conditions
\begin{equation}\label{2.12A}
\begin{split}
\gamma_1:=\sum_{i=1}^{N_1}|&\varphi_i\rangle\langle \varphi_i|+(\lambda_1-N_1)|\varphi_{N_1}\rangle\langle \varphi_{N_1}|,\ \   \varphi_i\in H^1(\R^3),\\
& (\varphi_i, \varphi_{i'})_{L^2}=\delta_{ii'},\ \ i, i'=1, 2, \cdots, N_1,
\end{split}
\end{equation}
\begin{equation}\label{2.12b}
\begin{split}
\gamma_2:=\sum_{j=1}^{N_2}|&\psi_j\rangle\langle\psi_j|+(\lambda-\lambda_1-N_2)|
\psi_{N_2}\rangle\langle\psi_{N_2}|,\ \   \psi_j\in H^1(\R^3),\\
  &(\psi_j, \psi_{j'})_{L^2}=\delta_{jj'},\ \  j, j'=1, 2, \cdots, N_2,
\end{split}\end{equation}
where $N_1,N_2\in\mathbb{N}^+$ are the smallest integers such that $\lambda_1\leq N_1$ and $ \lambda-\lambda_1\leq N_2$, respectively.
 For any fixed $\tau>0$, we define
$$
\psi_j^\tau(x):=\psi_j(x-\tau e_1),\ \  j=1,\cdots,N_2,\ \ e_1=(1,0,0),
$$
and consider the Gram matrix $G_\tau$ of the family $\varphi_1, \cdots,\ \varphi_{N_1},\  \psi_1^\tau, \cdots,  \psi^\tau_{N_2}$, i.e.,
\begin{equation}\label{2.11}
G_\tau:=
\left(
\begin{split}
\mathbb{I}_{N_1}\ \ \ \ &A_\tau\\
A_\tau^*\ \ \ \ & \mathbb{I}_{N_2}
\end{split}
\right), \ \   A_\tau=(a^\tau_{ij})_{{N_1}\times N_2},\ \  a^\tau_{ij}=(\varphi_i, \psi_j^\tau),
\end{equation}
where $	\mathbb{I}_{N_i}$ denotes the  ${N_i}$-order identity matrix.

Since
\begin{equation}\label{2.16b}
a_\tau:=\max\limits_{i, j}|(\varphi_i, \psi_j^\tau)|=o(1)\ \ \text{as}\ \  \tau\rightarrow\infty,
\end{equation}
it yields that   $G_\tau$ is positive definite for sufficiently large $\tau>0$. Hence,  the identity
\begin{equation}\label{2.16a}
\begin{split}
\mathbb{I}_{N_1+N_2}
=G_\tau^{-\frac{1}{2}}
\left(
\begin{split}
\mathbb{I}_{N_1}\ \ \ \ &A_\tau\\
A_\tau^*\ \ \ \ & \mathbb{I}_{N_2}
\end{split}
\right)
G_\tau^{-\frac{1}{2}}
\end{split}
\end{equation}
holds for sufficiently large $\tau>0$.
We next set
\begin{equation}\label{2.13}
(\tilde{\varphi}_1^\tau, \cdots,\ \tilde{\varphi}_{N_1}^\tau,\  \tilde{\psi}_1^\tau, \cdots, \tilde{\psi}^\tau_{N_2}):=(\varphi_1, \cdots,\ \varphi_{N_1},\  \psi_1^\tau, \cdots, \psi^\tau_{N_2})G_\tau^{-\frac{1}{2}},\ \ \tau>0,
\end{equation}
and
\begin{equation}\label{2.14}
\begin{split}
\gamma_\tau:=&\sum_{i=1}^{N_1}|\tilde{\varphi}_i^\tau\rangle\langle\tilde{\varphi}_i^\tau|+(\lambda_1-{N_1})|\tilde{\varphi}_{N_1}^\tau\rangle\langle\tilde{\varphi}_{N_1}^\tau|\\
&+\sum_{j=1}^{N_2}|\tilde{\psi}^\tau_j\rangle\langle\tilde{\psi}^\tau_j|+
(\lambda-\lambda_1-N_2)|\tilde{\psi}^\tau_{N_2}\rangle\langle\tilde{\psi}^\tau_{N_2}|,\ \ \tau>0.
\end{split}
\end{equation}
It then follows from \eqref{2.16a}  that the system $(\tilde{\varphi}_1^\tau,\cdots,\tilde{\varphi}_{N_1}^\tau, \tilde{\psi}^\tau_1,\cdots,\tilde{\psi}^\tau_{N_2})$ is an orthonormal family in $L^2(\R^3)$ for sufficiently large $\tau>0$, and hence $\gamma_\tau\in \mathcal{K}_{\lambda}$ defined by \eqref{set} holds for sufficiently large $\tau>0$.

We now calculate $\mathcal{E}_a(\gamma_{\tau})$ as $\tau \to\infty$. Since it follows from (\ref{2.11})  and \eqref{2.16b} that
\begin{equation}\label{tay}
\begin{split}
G_\tau^{-\frac{1}{2}}=
\left(\begin{split}
\mathbb{I}_{N_1}\ \ \ \ \ \ &0\\
0\ \ \ \ \ \ \ &\mathbb{I}_{N_2}
\end{split}
\right)
-\frac{1}{2}\left(
\begin{split}
0\ \ \ \ \ \ &A_\tau\\
A_\tau^*\ \ \ \ &0
\end{split}
\right)
+O(a_\tau^2)\ \ \ \mathrm{as}\ \ \tau\to\infty,
\end{split}
\end{equation}
one
can calculate from  \eqref{2.13} and \eqref{2.14}  that
\begin{equation}\label{2.15}
\begin{split}
\gamma_\tau=&\ \gamma_1+\gamma^\tau_2-\sum_{i=1}^{N_1}\sum_{j=1}^{N_2}a^\tau_{ij}\left(|\varphi_i\rangle\langle \psi_j^\tau|+|\psi_j^\tau\rangle\langle \varphi _i|\right)\\
&-\frac{1}{2}(\lambda_1-{N_1})\sum_{j=1}^{N_2}a^\tau_{N_1j}\left(|\varphi_{N_1}\rangle\langle \psi_j^\tau|+|\psi_j^\tau\rangle\langle \varphi _{N_1}|\right)\\
&-\frac{1}{2}(\lambda-\lambda_1-{N_2})\sum_{i=1}^{{N_1}}a^\tau_{iN_2}\left(|\varphi_i\rangle\langle \psi_{N_2}^\tau|+|\psi_{N_2}^\tau\rangle\langle \varphi _i|\right)\\
&+O(a_\tau^2) \ \  \, \mbox{as} \ \ \tau\to\infty,
\end{split}
\end{equation}
where  $\gamma_1$ is as in (\ref{2.12A}), and $\gamma^\tau_2=\sum_{j=1}^{N_2}|\psi_j^\tau\rangle\langle\psi_j^\tau|+(\lambda-\lambda_1-{N_2})|\psi^\tau_{N_2}\rangle\langle\psi^\tau_{N_2}|$.
We thus deduce from   \eqref{2.16b} that
\begin{equation}\label{2.17}
\mathrm{Tr}(-\Delta\gamma_\tau)=\mathrm{Tr}(-\Delta\gamma_1)+\mathrm{Tr}(-\Delta\gamma_2)+o(1)\ \ \text{as } \ \tau\to \infty,
\end{equation}
and
\begin{equation}\label{2.17a}
\int_{\R^3}\big|\rho_{\gamma_\tau}-\rho_{\gamma_1}-\rho_{\gamma_2}(x-\tau e_1)\big|dx=o(1)\ \ \text{as } \ \tau\to \infty,
\end{equation}
where $\rho_{\gamma_2}(x-\tau e_1)=\rho_{\gamma^\tau_2}(x)$.
Recall (cf. \cite{ho}) the following Hoffmann-Ostenhof inequality
\begin{equation}\label{ho}
\mathrm{Tr}(-\Delta \gamma_\tau)\geq\int_{\R^3}|\nabla\sqrt{\rho_{\gamma_\tau}}|^2dx.
\end{equation}
Applying  Sobolev's embedding theorem,
we then get from \eqref{2.17} that $\{\rho_{\gamma_\tau}\}$ is bounded uniformly in $L^{r}(\R^3)$ for all $r\in[1,3]$ as $\tau\to\infty$. Combining this with \eqref{2.17a}, one can deduce from the interpolation inequality that
$$
\rho_{\gamma_\tau}(x)-\rho_{\gamma_1}(x)-\rho_{\gamma_2}(x-\tau e_1)\rightarrow0\ \ \ \mathrm{strongly\ in} \ \ L^{r}(\R^3)\ \ \text{as}\ \ \tau\to\infty, \ \ \ r\in[1,3),
$$
which further implies from (\ref{v}) that
\begin{equation}\label{2.18}
\begin{split}
\lim\limits_{\tau\to\infty}\int_{\R^3}V(x)\rho_{\gamma_\tau}dx
&=\lim\limits_{\tau\to\infty} \int_{\R^3}V(x)\Big(\rho_{\gamma_1}(x)+\rho_{\gamma_2}(x-\tau e_1)\Big)dx\\
&=\int_{\R^3}V(x)\rho_{\gamma_1}dx,
\end{split}
\end{equation}
and
\begin{equation}\label{2.19}
\begin{split}
\lim\limits_{\tau\to\infty}\int_{\R^3}\rho_{\gamma_\tau}^{\frac{5}{3}}dx
&=\lim\limits_{\tau\to\infty} \int_{\R^3}\Big(\rho_{\gamma_1}(x)+\rho_{\gamma_2}(x-\tau e_1)\Big)^{\frac{5}{3}}dx\\
&=\int_{\R^3}\Big(\rho_{\gamma_1}^{\frac{5}{3}}+\rho_{\gamma_2}^{\frac{5}{3}}\Big)dx.
\end{split}\end{equation}

Applying Lemma \ref{lem:2.0}, we now conclude from \eqref{2.17}, \eqref{2.18} and \eqref{2.19} that
\begin{equation}
\begin{split}
E_a(\lambda)&=	 \inf\limits_{\gamma'\in\mathcal{K}_{\lambda}}\mathcal{E}_a(\gamma')\leq\lim\limits_{\tau\to\infty}\mathcal{E}_a(\gamma_{\tau})\\
&= \mathcal{E}_a(\gamma_1)+\mathrm{Tr}(-\Delta\gamma_2)-a\int_{\R^3}\rho_{\gamma_2}^{\frac{5}{3}}dx.
\end{split}
\end{equation}
Since $\gamma_1$ and $\gamma_2$ are arbitrary, the above inequality further implies that
\begin{align}\label{2.5a}
&E_a(\lambda)\nonumber\\
\leq& E_a(\lambda_1)
+\inf\Big\{ \mathrm{Tr}(-\Delta\gamma)-a\int_{\R^3}\rho_{\gamma}^{\frac{5}{3}}dx:\, \gamma=\sum_{i=1}^{N_2}|u_i\rangle\langle u_i|+(\lambda-\lambda_1-N_2)|u_{N_2}\rangle\langle u_{N_2}|,\Big.\nonumber\\
&\ \ \ \ \ \ \ \ \ \ \ \ \ \ \ \  \ \ \ u_i\in H^1(\R^3),\  (u_i,u_j)_{L^2}=\delta_{ij},\  i,j=1,\cdots,N_2\Big\}\nonumber\\
:=&E_a(\lambda_1)+E_a^\infty(\lambda-\lambda_1)
\end{align}
holds for any $0<\lambda_1<\lambda\leq N$.

For fixed $0\leq a\leq a^*_N$,  it then follows from \eqref{k} that $	E_a^\infty(\lambda-\lambda_1)\geq 0$. Similar to (\ref{2.3M}) and \eqref{2.3}, where $\gamma_t$ is replaced by $(\gamma_2)_t$, we obtain that
$$
E_a^\infty(\lambda-\lambda_1)\leq\lim\limits_{t\rightarrow0}t^2\Big(\mathrm{Tr}(-\Delta\gamma_2)
-a\int_{\R^3}\rho_{\gamma_2}^{\frac{5}{3}}dx\Big)=0,\ \ 0<\lambda_1<\lambda\leq N,
$$
and hence $E_a^\infty(\lambda-\lambda_1)=0$. Together with \eqref{2.5a}, this shows that if $0\leq a\leq a^*_N$, then $E_a(\lambda)\leq E_a(\lambda_1)$ holds for any  $0<\lambda_1<\lambda\leq N$. Therefore,  if $0\leq a\leq a^*_N$, then $ E_a(\lambda)$ is decreasing in $\lambda\in(0,N]$.

3. Following (\ref{form}), let $\gamma^{(N)}=\sum_{i=1}^{R_N}|Q_i\rangle\langle Q_i|$  be a minimizer of $a^*_N$, where $N\geq R_N\in\mathbb{N}^+$, and the system $(Q_1, \cdots,Q_{R_N})$ satisfies  $(Q_i,Q_j)=\delta_{ij}$. Since the above analysis gives that  $E_{ a^*_N}(\lambda)$ is decreasing in $\lambda \in (0,N]$, we have
\begin{align*}
E_{ a^*_N}(N)&\leq E_{ a^*_N}(R_N)\leq\lim\limits_{t\rightarrow\infty}\mathcal{E}_{ a^*_N}\big(\gamma^{(N)}_t\big)\\
&\leq-\lim\limits_{t\rightarrow\infty}t\int_{\R^3}|x|^{-1}\rho_{\gamma^{(N)}}dx=-\infty,
\end{align*}
where
$$
\gamma_t^{(N)}:=\sum_{i=1}^{R_N}t^3\big|Q_i\big(t(\cdot-y_1)\big)\rangle\langle Q_i\big(t(\cdot-y_1)\big)\big|,
$$
and $y_1\in \R^3$ is given in \eqref{v}.
By the definition of $E_{ a}(N)$, we then deduce from above that
$$E_{ a}(N)\leq E_{ a^*_N}(N)\leq-\infty,\ \  \ \forall \ a\geq a^*_N,$$
which completes the proof of Lemma \ref{lem2.1}.
\qed

\begin{rem}\label{2:rem:1}
Consider any fixed $N\geq2$, so that $0<a_N^*< a_1^*$ holds in view of \cite{ii}: If $a>0$ satisfies $(a_N^*<)a<a_1^*$, then it follows from Lemma \ref{lem2.1} $(1)$ that $-\infty<E_a(\lambda)<0$ holds for $\lambda=1<N$; If $a>0$ satisfies $a> a_1^*>a_N^*$, then it follows from Lemma \ref{lem2.1} $(3)$ that $E_a(\lambda)=-\infty$ holds for $\lambda=1<N$. On the other hand, consider $N=3$ and let $a>0$ be fixed and satisfy $a_3^*\leq a_2^*<a<a_1^*$: We obtain from Lemma \ref{lem2.1} $(1)$ that $-\infty<E_a(\lambda)<0$ holds for any $0<\lambda\leq1<N$; it however follows from Lemma \ref{lem2.1} $(3)$ that $E_a(\lambda)=-\infty$ holds for $\lambda=2<N$. These examples show that for any given $N\in \mathbb{N}^+$, if $a\geq a^*_N$ and  $\lambda \in (0, N)$, then $E_a(\lambda )$ can be either bounded or unbounded, which depends on the exact values of $a$ and $\lambda \in (0, N)$.
\end{rem}

Applying Lemmas \ref{lem:2.0} and \ref{lem2.1}, one can obtain the following analytical properties of minimizers for $E_a(\lambda)$.

\begin{lem}\label{lem2.2} For any fixed $N\in\mathbb{N}^+$, let $E_a(\lambda)$ be defined by \eqref{2.0}, where $a\in\big(0, a^*_N\big)$, $\lambda\in(0,N]$, and $V(x)<0$ is as in \eqref{v}. Suppose $\gamma=\sum_{i=1}^{N'}|u_i\rangle\langle u_i|+(\lambda-N')|u_{N'}\rangle\langle u_{N'}|$ is a minimizer of $E_a(\lambda)$, where $N'$ denotes the smallest integer such that $\lambda\leq N'$. Then we have
\begin{enumerate}
\item  $(u_1,\cdots,  u_{N'})$ is a ground state of the following system
\begin{equation*}\label{6}
\Big[-\Delta+V(x)-\frac{5}{3}a\Big(\sum_{j=1}^{N'}u_j^2+(\lambda-N')u_{N'}^2\Big)^{\frac{2}{3}}\Big]u_i=\mu_i u_i\ \ \ in\ \,\R^3,\ \ i=1,\cdots,N',
\end{equation*}	
where $\mu_1<\mu_2\leq\cdots\leq\mu_{N'}<0$ are the $N'$ first eigenvalues, counted with multiplicity, of the operator
\[H_V :=-\Delta+V(x)-\frac{5}{3}a\Big(\sum_{j=1}^{N'}u_j^2+(\lambda-N')u_{N'}^2\Big)^{\frac{2}{3}}  \ \ \  \,\ in\,\ \R^3.\]

\item $(u_1,\cdots,  u_{N'})$ decays exponentially in the sense that
\begin{gather}\label{2.6}
C^{-1}(1+|x|)^{-1}e^{-\sqrt{|\mu_1|}|x|}\leq u_1(x)\leq C(1+|x|)^{\frac{K}{\sqrt{|\mu_1|}}-1}e^{-\sqrt{|\mu_1|}|x|}\ \ in\ \,\R^3,
\end{gather}	
and
\begin{gather}\label{2.7}
|u_i(x)|\leq C(1+|x|)^{\frac{K}{\sqrt{|\mu_i|}}-1}e^{-\sqrt{|\mu_i|}|x|} \ \ in\ \, \R^3,\ \ i=2,\cdots, N',
\end{gather}
where the constant $K > 0$ is as in \eqref{v}, and $C>0$ depends on $\|\rho_{\gamma}\|_{L^{3}(\R^3)}$.
\end{enumerate}
\end{lem}

Since the proof of Lemma \ref{lem2.2} is similar to that of \cite[Lemma 2.3]{me}, we omit the
details for simplicity.

\subsection{Proof of Theorem \ref{th1}}

This subsection is devoted to the proof of Theorem \ref{th1}, for which we shall make full use of Lemmas \ref{lem2.1} and \ref{lem2.2}.

\vspace{.15cm}
\noindent\textbf{Proof of Theorem \ref{th1}.} In view of Lemmas \ref{lem2.1} and \ref{lem2.2}, we only need to prove the existence of minimizers for $E_a(N)$, where $N\in\mathbb{N}^+$ and $0<a<a^*_N$.

Consider any fixed $N\in\mathbb{N}^+$ and $0<a<a^*_N$. It follows from  Lemma \ref{lem2.1} that $E_a(N)$ is finite. 
Let $\{\gamma_n\}$ be a minimizing sequence of $E_a(N)$ with
$\gamma_n=\sum_{i=1}^{N}|u_i^n\rangle\langle u_i^n|$, where $(u_i^n, u_j^n)_{L^2}=\delta_{ij},\ \  i, j=1, 2, \cdots, N$.  The inequality \eqref{2.2} yields that the sequence $$
\big\{\mathrm{Tr}(-\Delta\gamma_n)\big\}=\Big\{\sum_{i=1}^N\int_{\R^3}|\nabla u_i^n|^2dx\Big\}
$$
is bounded uniformly in $n$, and thus $\{u_i^n\}_{n=1}^{\infty}$ is bounded uniformly in $H^1(\R^3)$ for all $i=1,\cdots, N$. Hence, one can assume that, up to a subsequence if necessary, there exists $u_i \in H^1(\R^3)$ such that
\begin{eqnarray}\label{2.8}
u_i^n\rightharpoonup  u_i\ \ \mathrm{weakly\ in}\, \ H^1(\R^3)\ \ \text{as}\ \ n\to\infty,\ \   i=1,\cdots,N,
\end{eqnarray}
and
\begin{eqnarray}\label{2.8a}
\ \rho_{\gamma_n}\!=\!\sum_{i=1}^{N}|u_i^n|^2\rightarrow \rho_\gamma:=\sum_{i=1}^{N}u_i^2\ \ \mathrm{strongly\ in} \,\ L^r_{loc}(\R^3)\ \ \text{as}\ \ n\to\infty, \ \ 1\leq r<3,
\end{eqnarray}
where $\gamma=\sum_{i=1}^{N}|u_i\rangle\langle u_i|$.

We first claim that if
\begin{equation}\label{assume}
\rho_{\gamma_n}\rightarrow \rho_\gamma\ \ \mathrm{ strongly\  in}\  L^1(\R^3)\ \ \mathrm{as}\ \ n\to\infty,
\end{equation}
then $\gamma$ is a minimizer of $E_a(N)$. Actually, using the weak lower semicontinuity, together with the fact that $(u_i^n, u_j^n)=\delta_{ij}$, $i,j=1,\cdots,N$,  we deduce from \eqref{assume} that 
\begin{equation}\label{2.15a}
u_i^n\rightarrow u_i\ \ \mathrm{strongly\ in} \ L^2(\R^3)\ \mathrm{as} \ n\to\infty,\ \ \mathrm{and}\ \ (u_i,u_j)=\delta_{ij},\ i,j=1,\cdots,N,
\end{equation}
which yields that
$$
\mathcal{E}_a(\gamma)\geq E_a(N), \ \ \text{where}\  \gamma=\sum_{i=1}^{N}|u_i\rangle\langle u_i|.
$$
Using the interpolation inequality and the boundedness of $\{\rho_{\gamma_n}\}$ in $L^{3}(\R^3)$, we derive from \eqref{assume} that
$$
\rho_{\gamma_n}\rightarrow \rho_\gamma\ \ \mathrm{ strongly\  in}\,\  L^r(\R^3)\ \mathrm{as} \ n\to\infty,\ 1\leq r<3.
$$
Therefore, we have
\begin{align*}
E_a(N)&=\liminf\limits_{n\to\infty}\mathcal{E}_a(\gamma_n)\\
&\geq\mathrm{Tr}(-\Delta\gamma)+\int_{\R^3}V(x)\rho_\gamma dx-a\int_{\R^3}\rho_\gamma^{\frac{5}{3}}dx\\
&=\mathcal{E}_a(\gamma)\geq E_a(N),
\end{align*}
which implies that $\gamma$ is a minimizer of $E_a(N)$.

As a consequence,  in order to prove Theorem \ref{th1}, the rest is to prove (\ref{assume}). Applying the Br$\acute{e}$zis-Lieb Lemma (cf. \cite{minmax}), we note from \eqref{2.8} that if  $\int_{\R^3}\rho_\gamma dx=N$, then (\ref{assume}) holds true. Therefore, the rest proof of Theorem \ref{th1} is to prove
by two steps that the case $\lambda:=\int_{\R^3}\rho_\gamma dx\in[0, N)$ cannot occur. We shall denote $\rho_{n}:=\rho_{\gamma_n}$ for convenience.

{\em Step 1.} We first prove that the case  $\lambda:=\int_{\R^3}\rho_\gamma dx=0$ cannot occur. On the contrary, suppose $\lambda=0$. It then follows from \eqref{2.8a} that
$$
\lim\limits_{n\to\infty}\int_{\R^3}V(x)\rho_{n}dx
=-\sum_{k=1}^K\lim\limits_{n\to\infty}\int_{\R^3}|x-y_k|^{-1}\rho_{n}dx=0.
$$
We thus get from (\ref{k}) that
\begin{align*}
E_a(N)&=\lim\limits_{n\to\infty}\mathcal{E}_a(\gamma_n)\\
&=\lim\limits_{n\to\infty}\Big[\mathrm{Tr}(-\Delta \gamma_n)-a\int_{\R^3}\rho_{n}^{\frac{5}{3}}dx\Big]+\lim\limits_{n\to\infty}\int_{\R^3}V(x)\rho_{n}dx\\
&\geq\Big(1-\frac{a}{a^*_N}\Big)\liminf\limits_{n\to\infty}\mathrm{Tr}(-\Delta \gamma_n)\geq0,
\end{align*}
which however contradicts with Lemma \ref{lem2.1} (1). Thus, the case $\lambda=0$ cannot occur.

{\em Step 2.} We next  prove that the case
$0<\lambda:=\int_{\R^3}\rho_\gamma dx<N$ cannot occur, either. By contradiction, suppose $0<\int_{\R^3}\rho_\gamma dx=\lambda<N$. By an adaptation of the classical dichotomy result (cf. \cite[Section 3.3]{begain}), up to a subsequence of $\{\rho_n\}$  if necessary, then there exists a sequence $\{R_n\}$ with $R_n\rightarrow\infty$ as $n\to\infty$ such that
\begin{equation}\label{dic}
0<\lim\limits_{n\to\infty}\int_{|x|\leq R_n}\rho_ndx=\int_{\R^3}\rho_\gamma dx<N,\ \ \ \lim\limits_{n\to\infty}\int_{R_n\leq|x|\leq 6R_n}\rho_ndx=0.
\end{equation}
Choose $\chi\in C_0^\infty(\R^3, [0,1])$ satisfying $\chi(x)=1$ for $|x|\leq1$ and $\chi(x)=0$ for $|x|\geq2$, and define 
$\chi_{R_n}(x):=\chi(x/R_n)$, $\eta_{R_n}(x):=\sqrt{1-\chi_{R_n}^2(x)}$,
\begin{equation}\label{2.36a}
u_i^{1n}:=\chi_{R_n}u_i^n,\ \    u_i^{2n}:=\eta_{R_n}u_i^n,\ \   i=1,\cdots,N,
\end{equation}
and
\begin{equation}\label{2.37}
\gamma_{1n}:=\sum_{i=1}^{N}|u_i^{1n}\rangle \langle u_i^{1n}|, \ \    \gamma_{2n}:=\sum_{i=1}^{N}|u_i^{2n}\rangle \langle u_i^{2n}|.
\end{equation}

We now follow above to estimate the energy $\mathcal{E}_a(\gamma_n)$ as $n\to\infty$. It is easy to verify from \eqref{2.8} and \eqref{2.36a} that \begin{eqnarray}\label{2.38}
	u_i^{1n}\rightharpoonup  u_i\ \ \mathrm{weakly\ in}\, \ H^1(\R^3)\ \ \text{as}\ \ n\to\infty,\ \   i=1,\cdots,N.
\end{eqnarray}
Using the weak lower semicontinuity of norm,  
together with the Br$\acute{e}$zis-Lieb Lemma (cf. \cite{minmax}), one can deduce from \eqref{dic}  and \eqref{2.38} that
\begin{equation}\label{2.32a}
\liminf\limits_{n\rightarrow\infty}\mathrm{Tr}(-\Delta\gamma_{1n})\geq \mathrm{Tr}(-\Delta\gamma),
\end{equation}
and
\begin{equation}\label{2.35}
\rho_{1n}:=\rho_{\gamma_{1n}}\rightarrow\rho_\gamma\ \ \mathrm{strongly\ in}\ \ L^1(\R^3)\ \ \mathrm{as}\ \ n\to\infty.
\end{equation}
Moreover, we have
\begin{equation}\label{2.35b}
\rho_n=\chi_{R_n}^2\rho_{n}+\eta_{R_n}^2\chi_{3R_n}^2\rho_n+\eta_{3R_n}^2\rho_n,
\end{equation}
and
\begin{equation}\label{2.36}
\eta_{R_n}^2\chi_{3R_n}^2\rho_n\rightarrow0\ \ \ \mathrm{strongly\ in}\ \ L^1(\R^3)\ \ \mathrm{as} \ \ n\to\infty,
\end{equation}
due to the estimate \eqref{dic}.
Following the uniform boundedness of $\{\rho_n\}$ in $L^{1}(\R^3)\cap L^{3}(\R^3)$, we derive from \eqref{2.35} and \eqref{2.36} that
\begin{equation}\label{2.32b}
\rho_{1n}:=\rho_{\gamma_{1n}}\rightarrow\rho_\gamma,\ \ \   \eta_{R_n}^2\chi_{3R_n}^2\rho_n\rightarrow0\ \ \mathrm{strongly\ in}\ \ L^r(\R^3)\ \ \text{as}\ \ n\to\infty,\ \ \, r\in[1,3).
\end{equation}
We thus obtain from \eqref{2.35b} and \eqref{2.32b} that for $\rho_{2n}:=\rho_{\gamma_{2n}}$,
\begin{equation}\label{14}
\begin{split}
\int_{\R^3}V(x)\rho_{n}dx
&=\int_{\R^3}V(x)\rho_{1n}dx+\int_{\R^3}V(x)\rho_{2n}dx\\
&=\int_{\R^3}V(x)\rho_{\gamma}dx+o(1) \ \ \ \mbox{as}\ \ n\to\infty,
\end{split}
\end{equation}
and
\begin{equation}\label{16}
\begin{split}
\int_{\R^3}\rho_{n}^{\frac{5}{3}}dx
=&\int_{\R^3}\left(\chi_{R_n}^2\rho_{n}
+\eta_{3R_n}^2\rho_{n}\right)^{\frac{5}{3}}dx+o(1) \\
=& \int_{\R^3}\Big[\left(\chi_{R_n}^2\rho_{n}\right)^{\frac{5}{3}}
+\left(\eta_{3R_n}^2\rho_{n}\right)^{\frac{5}{3}}\Big]dx+o(1)\\
=& \int_{\R^3}\Big[\left(\chi_{R_n}^2\rho_{n}\right)^{\frac{5}{3}}
+\left(\eta_{R_n}^2\chi_{3R_n}^2\rho_n+\eta_{3R_n}^2\rho_{n}\right)^{\frac{5}{3}}\Big]dx+o(1)\\
=&\int_{\R^3}\Big(\rho_{\gamma}^{\frac{5}{3}}+\rho_{2n}^{\frac{5}{3}}\Big)dx+o(1)\ \   \mathrm{as}\ \, n\to\infty.
\end{split}
\end{equation}
Since it follows from \cite[Theorem 3.1]{ims} that $$-\Delta=\chi_{R_n}(-\Delta)\chi_{R_n}+\eta_{R_n}(-\Delta)\eta_{R_n}-|\nabla\chi_{R_n}|^2-|\nabla\eta_{R_n}|^2,$$
we have
\begin{equation}\label{2.30}
\begin{split}
\mathrm{Tr}(-\Delta\gamma_n)
=&\ \mathrm{Tr}(-\Delta\gamma_{1n})+\mathrm{Tr}(-\Delta\gamma_{2n})
-\int_{\R^3}(|\nabla\chi_{R_n}|^2+|\nabla\eta_{R_n}|^2)\rho_ndx\\
\geq&\ \mathrm{Tr}(-\Delta\gamma_{1n})+\mathrm{Tr}(-\Delta\gamma_{2n})
-CR_n^{-2}N,
\end{split}
\end{equation}
where $C>0$ is independent of $n>0$.  We therefore conclude from \eqref{2.32a} and \eqref{14}--\eqref{2.30} that
\begin{equation}\label{2.4}
\begin{split}
E_a(N)=\lim\limits_{n\to\infty}\mathcal{E}_a(\gamma_n)
&\geq\mathcal{E}_a(\gamma)
+\liminf\limits_{n\to\infty}\Big(\mathrm{Tr}(-\Delta\gamma_{2n})-a\int_{\R^3}\rho_{2n}^{\frac{5}{3}}dx\Big).
\end{split}
\end{equation}

Note from (\ref{2.8a}) that $\gamma=\sum_{i=1}^{N}|u_i\rangle\langle u_i|\in\mathcal{K}_{\lambda}$, where $\lambda=\int_{\R^3}\rho_\gamma dx$ and  $\mathcal{K}_{\lambda}$ is defined by Lemma \ref{lem:2.0}. We thus get from Lemma \ref{lem:2.0} that
\begin{equation}\label{2.45}
\mathcal{E}_a(\gamma)\geq \inf\limits_{\gamma'\in\mathcal{K}_\lambda}\mathcal{E}_a(\gamma')=E_a(\lambda).
\end{equation}
Applying Lemma \ref{lem2.1} (2), we further obtain from \eqref{2.4} and \eqref{2.45} that
\begin{equation}\label{2.12}
\begin{split}
E_a(\lambda)\geq E_a(N)
&\geq \mathcal{E}_a(\gamma)
  +\liminf\limits_{n\to\infty}\Big(\mathrm{Tr}(-\Delta\gamma_{2n})-a\int_{\R^3}\rho_{2n}^{\frac{5}{3}}dx\Big)\\
&\geq \inf\limits_{\gamma'\in\mathcal{K}_\lambda}\mathcal{E}_a(\gamma')
  +\liminf\limits_{n\to\infty}\Big(\mathrm{Tr}(-\Delta\gamma_{2n})-a\int_{\R^3}\rho_{2n}^{\frac{5}{3}}dx\Big)\\
&=E_a(\lambda)+ \liminf\limits_{n\to\infty}
 \Big(\mathrm{Tr}(-\Delta\gamma_{2n})-a\int_{\R^3}\rho_{2n}^{\frac{5}{3}}dx\Big)\\
&\geq E_a(\lambda)+\liminf\limits_{n\to\infty} \Big(1-\frac{a\|\gamma_{2n}\|^{\frac{2}{3}}}{a^*_N}\Big)\mathrm{Tr}(-\Delta \gamma_{2n})\\[1mm]
&\geq E_a(\lambda)+\Big(1-\frac{a}{a^*_N}\Big)\liminf\limits_{n\to\infty} \mathrm{Tr}(-\Delta \gamma_{2n}),
\end{split}
\end{equation}
where the last inequality follows from the fact that $\|\gamma_{2n}\|\leq\|\gamma_n\|=1$.
Thus, if $\liminf\limits_{n\to\infty}\mathrm{Tr}(-\Delta \gamma_{2n})>0$, then we get a contradiction from (\ref{2.12}), and thus the case
$0<\lambda:=\int_{\R^3}\rho_\gamma dx<N$ cannot occur, which therefore completes the proof of Theorem \ref{th1}.

If $\liminf\limits_{n\to\infty}\mathrm{Tr}(-\Delta \gamma_{2n})=0$, we derive from \eqref{2.12} that
\begin{equation}\label{2.9}
E_a(\lambda)=E_a(N),
\end{equation}
and $\gamma$ is a minimizer of $\inf\limits_{\gamma'\in\mathcal{K}_\lambda}\mathcal{E}_a(\gamma')$. This further implies from Lemma \ref{lem:2.0} that $E_a(\lambda)$ possesses minimizers, where $0<\lambda:=\int_{\R^3}\rho_\gamma dx<N$. We next consider two different cases.

{\em Case 1: $N=1$.} For the case $N=1$, since $\int_{\R^{3}}\rho_{\gamma}dx=\int_{\R^{3}}u_1^2dx=\lambda$,
we deduce from \eqref{2.9} that for $\varphi:=\lambda^{-\frac{1}{2}}u_1$,
\begin{align*}
E_a(1)=E_a(\lambda)=&\inf\limits_{\gamma'\in\mathcal{K}_\lambda}\mathcal{E}_a(\gamma')=\mathcal{E}_a(\gamma)\\
=&\int_{\R^{3}}\Big(|\nabla u_1|^2+V(x)u_1^2-au_1^{\frac{10}{3}}\Big)dx\\
=&\lambda\int_{\R^{3}}\Big(|\nabla \varphi|^2+V(x)\varphi^2-a\lambda^{\frac{2}{3}}\varphi^{\frac{10}{3}}\Big)dx\\
=&\lambda\mathcal{E}_a(|\varphi\rangle\langle \varphi|)+a\lambda\big(1-\lambda^{\frac{2}{3}}\big)\int_{\R^{3}}\varphi^{\frac{10}{3}}dx\\
\geq &\lambda E_a(1)+a\lambda\big(1-\lambda^{\frac{2}{3}}\big)\int_{\R^{3}}\varphi^{\frac{10}{3}}dx.
\end{align*}
If $0<\lambda:=\int_{\R^3}\rho_\gamma dx<1=N$, then one has $E_a(1)>\lambda E_a(1)$, which however contradicts with the fact that $E_a(1)<0$. Therefore, if $N=1$, then $0<\lambda:=\int_{\R^3}\rho_\gamma dx<N$ cannot occur.

{\em Case 2: $N\ge 2$.} As for the case $2\leq N\in\mathbb{N}^+$, there exists an integer $N'\in [1,  N]$ such that $\lambda\in[N'-1,N')$. Let
\begin{equation}\label{2.38A}
\gamma_1:=\sum_{i=1}^{N'}|\varphi_i\rangle\langle\varphi_i|+(\lambda-{N'})|\varphi_{N'}\rangle\langle\varphi_{N'}|
\end{equation}
be a minimizer of $E_a(\lambda)$, where $(\varphi_i, \varphi_j)=\delta_{ij}$. Consider
$$
\gamma_2:=\sum_{j=1}^{N-{N'}+1}|\psi_j\rangle\langle\psi_j|+({N'}-\lambda-1)|\psi_{N-{N'}+1}\rangle\langle\psi_{N-{N'}+1}|,
$$
where the functions $\psi_1,\cdots,\psi_{N-{N'}+1}\in C_0^\infty(\R^3, [0, 1])$ satisfy  $(\psi_i,\psi_j)=\delta_{ij}$. 
Denote
\begin{equation}\label{2.32M}
\psi_j^\tau(x):=\tau^{-3/2}\psi_j(\tau^{-1} x), \ \ \mbox{where}\ \ \tau>0, \ \ j=1,\cdots,N-N'+1.
\end{equation}
Since  $\gamma_1$ is a minimizer of $E_a(\lambda)$,
it follows from  \eqref{2.38A} and Lemma \ref{lem2.2} that  there exist   constants $C >0$ and $ \theta>0$ such that
\begin{equation}\label{2.33B}
|\varphi_i(x)|\leq Ce^{-\theta|x|}\ \ \mbox{in}\ \ \R^3, \ \  i=1,\cdots, {N'},
\end{equation}
which yields from (\ref{2.32M}) that
\begin{equation}\label{2.33}
0\leq a_\tau:=\max\limits_{i,j}\big\{|(\varphi_i, \psi_j^\tau)|\big\}\leq C_1\tau^{-\frac{3}{2}}\ \ \ \text{as}\ \  \tau\rightarrow\infty,
\end{equation}
where $C_1>0$ is independent of $\tau>0$.
We thus can define the same operator $\gamma_\tau\in \mathcal{K}_{N}$ as in \eqref{2.14}, where $\tau>0$ is sufficiently large.

Similar to \eqref{2.15}, one can get that
\begin{align}\label{2.32}
\gamma_\tau
=&\ \gamma_1+\gamma^\tau_2-\sum_{i=1}^{N'}\sum_{j=1}^{N-{N'}+1}a^\tau_{ij}\left(|\varphi_i\rangle\langle \psi_j^\tau|+|\psi_j^\tau\rangle\langle \varphi _i|\right)\nonumber\\
&-\frac{1}{2}(\lambda-{N'})\sum_{j=1}^{N-{N'}+1}a^\tau_{N'j}\left(|\varphi_{N'}\rangle\langle \psi_j^\tau|+|\psi_j^\tau\rangle\langle \varphi _{N'}|\right)\\
&-\frac{1}{2}({N'}-\lambda-1)\sum_{i=1}^{{N'}}a^\tau_{i,N-{N'}+1}\left(|\varphi_i\rangle\langle \psi_{N-{N'}+1}^\tau|+|\psi_{N-{N'}+1}^\tau\rangle\langle \varphi _i|\right)\nonumber+O(a_\tau^2) \nonumber\\
:=&\gamma_1+\gamma^\tau_2-\gamma_{A_\tau}\ \ \  \mbox{as} \ \ \tau\to\infty\nonumber,
\end{align}
where $a^\tau_{ij}:=(\varphi_i, \psi_j^\tau)$, $a_\tau:=\max\limits_{i,j}\big\{|(\varphi_i, \psi_j^\tau)|\big\}$
and
$$\gamma^\tau_2:=\sum_{j=1}^{N-{N'}+1}|\psi_j^\tau\rangle\langle\psi_j^\tau|
+({N'}-\lambda-1)|\psi^\tau_{N-{N'}+1}\rangle\langle\psi^\tau_{N-{N'}+1}|.$$

Applying Lemma \ref{lem2.2}, we thus deduce from \eqref{2.33B}--\eqref{2.32} that
\begin{equation}\label{2.20a}
\begin{split}
\mathrm{Tr}(-\Delta+V(x))\gamma_\tau
=&\ \mathrm{Tr}(-\Delta+V(x))(\gamma_1 +\gamma^\tau_2)+O(a_\tau^2)\\[1mm]
=&\ \mathrm{Tr}(-\Delta+V(x))\gamma_1+\tau^{-2}\mathrm{Tr} (-\Delta\gamma_2)\\
&-\tau^{-1}\sum_{k=1}^K\int_{\R^3}\big|x-\tau^{-1}y_k\big|^{-1}\rho_{\gamma_2}dx+O(a_\tau^2)\\
\leq&\ \mathrm{Tr}(-\Delta+V(x))\gamma_1-C_2\tau^{-1}
\ \ \mathrm{as}\ \ \tau\to\infty,
\end{split}
\end{equation}
and
\begin{align}\label{2.21a}
\int_{\R^3}\rho_{\gamma_\tau}^{5/3}dx
\geq&\int_{\R^3}\Big(\rho_{\gamma_1}-|\rho_{\gamma_{A_\tau}}|\Big)^{5/3}dx\nonumber\\
=&\int_{\R^3}\rho_{\gamma_1}^{5/3}dx
-\frac{5}{3}\big(1+o(1)\big)\int_{\R^3}\rho_{\gamma_1}^{2/3}|\rho_{\gamma_{A_\tau}}|dx\\
=&\int_{\R^3}\rho_{\gamma_1}^{5/3}dx-O(a^2_\tau)\ge \int_{\R^3}\rho_{\gamma_1}^{5/3}dx-C_2\tau^{-3}\ \ \mathrm{ as}\ \  \tau\rightarrow\infty,\nonumber
\end{align}
where $C_2>0$ is independent of $\tau>0$, and the operator $\gamma_{A_\tau}$ is defined in \eqref{2.32}. As a consequence, we conclude from \eqref{equ}, \eqref{2.20a} and \eqref{2.21a} that
\begin{equation}\label{2.57A}
E_a(N)=\inf\limits_{\gamma'\in\mathcal{K}_N}\mathcal{E}_a(\gamma')
\leq \mathcal{E}_a(\gamma_\tau)\leq E_a(\lambda)	-\frac{C_2}{2}\tau^{-1}<E_a(\lambda)\ \ \mathrm{as}\ \  \tau\rightarrow\infty,
\end{equation}
which however contradicts with the identity  \eqref{2.9}. This shows that if $N\ge 2$, then $0<\lambda:=\int_{\R^3}\rho_\gamma dx<N$ cannot occur, either. The proof of Theorem \ref{th1} is therefore complete.\qed

\section{$N=2$: Limiting Behavior of Minimizers as $a\nearrow a^*_2$}
In this section, we address the proof of Theorem 1.2 on the limiting concentration behavior of minimizers  for $E_a(N)$ with $N=2$ as $a\nearrow a^*_2$, where $a^*_2>0$ is defined by \eqref{k}. Towards this purpose, we shall employ the first three steps from the so-called blow-up analysis of many-body fermionic problems, which is described in Subsection 1.1.

Throughout the rest part of this paper, we follow  \cite[ Theorem 6 and  Proposition 11]{ii}  to suppose that
\begin{equation}\label{q}
\gamma^{(2)}=\sum_{i=1}^{2}|Q_i\rangle\langle Q_i|\ \ \mathrm{with}\ \ (Q_i, Q_j)=\delta_{ij}
\end{equation}
is an optimizer of $a^*_{2}$, where $Q_i$ satisfies
\begin{equation}\label{3.0}
Q_i\in C^\infty(\R^3),\ \ \,    \mathrm{ and}\ \ |\nabla Q_i(x)|,\ \  |Q_i(x)|= O(e^{-\sqrt{|\hat{\mu}_i|}|x|})\ \ \mathrm{as}\ |x|\to \infty,\ \ i=1,2,
\end{equation}
and $\hat{\mu}_i<0$ denotes the $i$th eigenvalue (counted with multiplicity) of $-\Delta-\frac{5}{3}a^*_2\Big(\sum_{j=1}^2Q_j^2\Big)^{2/3}$  in $\R^3$.

We start with the following energy estimates of $E_a(2)$ as $a\nearrow a_2^*$.

\begin{lem}\label{lem3.1}
Suppose $\gamma_a$ is a minimizer of $E_a(2)$ defined by \eqref{1.2}, where $0<a<a^*_2$. Then there exist some  constants $0<M_1<M_2$, $0<M'_1<M'_2$, $0<M''_1$ and $0<M'''_1$, independent of $0<a<a^*_2$, such that
\begin{gather}\label{3.3}
M_1\leq-\epsilon_aE_a(2)\leq M_2,\ \   M_1'\leq -\epsilon_a\int_{\R^3}V(x)\rho_{\gamma_a}dx\leq M'_2\, \ \ as\ \ a\nearrow a^*_2,
\end{gather}
and
\begin{gather}\label{3.4}
0\leq\epsilon_a^2\mathrm{Tr}(-\Delta\gamma_a)\leq  M''_1,\ \    0\leq\epsilon_a^2\int_{\R^3}\rho_{\gamma_a}^{\frac{5}{3}}dx\leq M'''_1 \ \   as\ \ a\nearrow a^*_2,
\end{gather}
where $\epsilon_a:=a^*_2-a>0$, and the potential $V(x)=-\sum_{k=1}^K|x-y_k|^{-1}<0$ is as in \eqref{v}.
\end{lem}

\noindent{\bf Proof.}  Take a cut-off function $\varphi\in C_{0}^{\infty}(\R^3, [0,1])$ satisfying $\varphi(x) =1$ for $|x|\leq1$ and $\varphi(x)=0$ for $|x|\geq2$.  Define for $\tau>0$,
\begin{equation}\label{3.1a}
Q_i^\tau(x):=A_i^\tau\tau^{\frac{3}{2}}\varphi(x-y_1)Q_i\big(\tau (x-y_1)\big),\ \  \  i=1,2,
\end{equation}
where $Q_i\in C^\infty(\R^3)$ and $y_1\in\R^3$ are given by \eqref{q} and \eqref{v}, respectively, and $A_i^\tau>0$ is chosen such that $\int_{\R^3}|Q_i^\tau(x)|^2dx=1,\ i=1, 2$.
The exponential decay of $Q_i$ in \eqref{3.0} then gives that
\begin{equation}\label{o1}
A_i^\tau=1+o(\tau^{-\infty})\ \ \ \ \mathrm{and}\ \ \ \
a_\tau:=\big(Q_1^\tau, Q_2^\tau\big)=o(\tau^{-\infty})\ \ \ \mathrm{as}\ \ \tau\to\infty,
\end{equation}
where $o(\tau^{-\infty})$ means $ \lim\limits_{\tau\rightarrow\infty}o(\tau^{-\infty})\tau^s=0$ for any $s\geq0$. This implies that   the following Gram matrix
\begin{equation}\label{3.6b}
\begin{split}
G_\tau:=
\left(
\begin{split}
&Q_1^\tau\\
& Q_2^\tau
\end{split}
\right)
\big(Q_1^\tau, Q_2^\tau\big)
=\begin{bmatrix}
1&\   \big(Q_1^\tau, Q_2^\tau\big)\\
\big(Q_2^\tau, Q_1^\tau\big)&\ 1
\end{bmatrix}
=
\begin{bmatrix}
1&\   a_\tau\\
a_\tau&\ 1
\end{bmatrix}
\end{split}
\end{equation}
is positive definite for sufficiently large $\tau>0$.

We now define for $\tau>0$,
\begin{equation}\label{d0a}
\big(\tilde{Q}_1^\tau, \tilde{Q}_2^\tau\big):=\big(Q_1^\tau, Q_2^\tau\big)G_\tau^{-\frac{1}{2}}.
\end{equation}
It then follows from \eqref{3.6b} that for  sufficiently large $\tau>0$,
\begin{equation}\label{d0}
\big(\tilde{Q}_i^\tau, \tilde{Q}_j^\tau\big)=\delta_{ij},\ \   i,j=1,2.
\end{equation}
Similar to \eqref{tay}, one can obtain from (\ref{3.6b}) the Taylor's expansion of $G_\tau$ as $\tau\to\infty$. Thus we also derive from \eqref{d0a} that
\begin{equation}\label{d0b}
	\big(\tilde{Q}_1^\tau, \tilde{Q}_2^\tau\big)=\big(Q_1^\tau, Q_2^\tau\big)-\frac{1}{2}a_\tau\big(Q_2^\tau, Q_1^\tau\big)+O(a_\tau^2)\ \ \, \mathrm{as}\ \, \tau\to\infty.
\end{equation}
Setting
$$
\tilde{\gamma}^{(2)}_\tau:=
\sum_{i=1}^{2}|\tilde{Q}_i^\tau\rangle\langle \tilde{Q}_i^\tau|,
$$
we can  compute from \eqref{o1}   and \eqref{d0b} that
\begin{align}\label{d1}
\mathcal{E}_a(\tilde{\gamma}_\tau^{(2)})
=&\mathrm{Tr}\big(-\Delta+V(x)\big) \tilde{\gamma}^{(2)}_\tau-a\int_{\R^3}\rho_{\tilde{\gamma}^{(2)}_\tau}^{\frac{5}{3}}dx\nonumber\\
=&\sum_{i=1}^2\int_{\R^3}|\nabla\tilde{Q}_i^\tau|^2dx+\sum_{i=1}^2\int_{\R^3}V(x)|\tilde{Q}_i^\tau|^2dx-a\int_{\R^3}\Big(\sum_{i=1}^2|\tilde{Q}_i^\tau|^2\Big)^{\frac{5}{3}}dx\nonumber\\
=&\sum_{i=1}^2\int_{\R^3}|\nabla Q_i^\tau|^2dx+\sum_{i=1}^2\int_{\R^3}V(x)|Q_i^\tau|^2dx-a\int_{\R^3}\Big(\sum_{i=1}^2|Q_i^\tau|^2\Big)^{\frac{5}{3}}dx\nonumber\\
&-2a_\tau\int_{\R^3}\nabla Q_1^\tau\cdot\nabla Q_2^\tau dx-2a_\tau\int_{\R^3}V(x)Q_1^\tau Q_2^\tau dx\\
&+\frac{10}{3}a a_\tau\int_{\R^3}\Big(\sum_{i=1}^2|Q_i^\tau|^2\Big)^{\frac{2}{3}}Q_1^\tau Q_2^\tau dx+O(a_\tau^{2})\nonumber\\
=&\sum_{i=1}^2\int_{\R^3}|\nabla Q_i^\tau|^2dx+\sum_{i=1}^2\int_{\R^3}V(x)|Q_i^\tau|^2dx-a\int_{\R^3}
\Big(\sum_{i=1}^2|Q_i^\tau|^2\Big)^{\frac{5}{3}}dx\nonumber\\[2mm]
&+o(\tau^{-\infty})\ \ \  \mathrm{as}\ \ \tau\to\infty,\nonumber
\end{align}
where the second identity follows from the orthonormality of  \eqref{d0}.

To estimate the right hand side of (\ref{d1}), we calculate from \eqref{3.0} and  \eqref{3.1a} that
\begin{align}\label{3.11}
\sum_{i=1}^2\int_{\R^3}|\nabla Q_i^\tau|^2dx
=&\sum_{i=1}^2|A_i^\tau|^2\tau^{3}\int_{\R^3}\Big|Q_i\big(\tau (x-y_1)\big)\nabla\varphi(x-y_1)\Big.\nonumber\\
&\ \ \ \ \ \ \ \ \ \ \ \ \ \ \ \ \ \ \ \ \ \ \Big.+\tau\varphi(x-y_1)\nabla Q_i\big(\tau (x-y_1)\big)\Big|^2dx\\
=&\tau^{5}\sum_{i=1}^2\int_{\R^3}\varphi^2(x-y_1)\big|\nabla Q_i\big(\tau (x-y_1)\big)\big|^2dx+o(\tau^{-\infty})\nonumber\\
=&\tau^{2}\sum_{i=1}^2\int_{\R^3}\big|\nabla Q_i(x)\big|^2dx+o(\tau^{-\infty})\ \ \  \mathrm{as}\ \ \tau\to\infty,\nonumber
\end{align}
and
\begin{align}\label{3.12}
\sum_{i=1}^2\int_{\R^3}V(x)|Q_i^\tau|^2dx
\leq& -\sum_{i=1}^2\int_{\R^3}|x-y_1|^{-1}|Q_i^\tau|^2dx\nonumber\\
=&-\tau\sum_{i=1}^2\int_{\R^3}|x|^{-1}|A_i^\tau|^2\varphi^2(\tau^{-1}x)Q_i^2(x)dx\\
=& -\tau\sum_{i=1}^2 \int_{\R^3}|x|^{-1}Q_i^2(x)dx+o(\tau^{-\infty})\ \ \ \mathrm{as}\ \ \tau\to\infty.\nonumber
\end{align}
As for the nonlinear term, one has
\begin{align}\label{3.13}
\int_{\R^3}\Big(\sum_{i=1}^2|Q_i^\tau|^2\Big)^{\frac{5}{3}}dx
 =&\tau^2\int_{\R^3}\varphi^{\frac{10}{3}}\big(\tau^{-1}x\big)\Big(\sum_{i=1}^2|A_i^\tau|^2Q_i^2(x)\Big)^{\frac{5}{3}}dx\nonumber\\
=&\tau^2\int_{\R^3}\Big(\sum_{i=1}^2|A_i^\tau|^2Q_i^2(x)\Big)^{\frac{5}{3}}dx\nonumber\\ &-\tau^2\int_{\R^3}\Big(\sum_{i=1}^2|A_i^\tau|^2Q_i^2(x)\Big)^{\frac{5}{3}}\Big[1-\varphi^{\frac{10}{3}}(\tau^{-1}x)\Big]dx\\
=&\tau^2\int_{\R^3}\Big(\sum_{i=1}^2|A_i^\tau|^2Q_i^2(x)\Big)^{\frac{5}{3}}dx+o(\tau^{-\infty})\nonumber\\
=&\tau^2\int_{\R^3}\Big(\sum_{i=1}^2Q_i^2(x)\Big)^{\frac{5}{3}}dx+o(\tau^{-\infty})\ \ \  \mathrm{as}\ \ \tau\to\infty.\nonumber
\end{align}
Consequently, we obtain from  \eqref{d1}--\eqref{3.13} that
\begin{align}\label{3.14}
\mathcal{E}_a(\tilde{\gamma}_\tau^{(2)})
\leq&\tau^{2}\sum_{i=1}^2\int_{\R^3}\big|\nabla Q_i(x)\big|^2dx
-\tau\sum_{i=1}^2 \int_{\R^3}|x|^{-1}Q_i^2(x)dx\nonumber\\
&-a\tau^2\int_{\R^3}\Big(\sum_{i=1}^2Q_i^2(x)\Big)^{\frac{5}{3}}dx+o(\tau^{-\infty})\\
=&\tau^{2}\mathrm{Tr}(-\Delta\gamma^{(2)})-\tau\int_{\R^3}|x|^{-1}\rho_{\gamma^{(2)}}dx-a\tau^2\int_{\R^3}\rho_{\gamma^{(2)}}^{\frac{5}{3}}dx+o(\tau^{-\infty})\nonumber\\
=&\tau^{2}\big(a_2^*-a\big)\int_{\R^3}\rho_{\gamma^{(2)}}^{\frac{5}{3}}dx-\tau\int_{\R^3}|x|^{-1}\rho_{\gamma^{(2)}}dx+o(\tau^{-\infty})\ \ \ \mathrm{as}\ \ \tau\to\infty,\nonumber
\end{align}
where the last identity  follows from the fact that  $\gamma^{(2)}$ given in \eqref{q} is an optimizer of $a^*_2$.

Taking
$$
\tau=t\epsilon_a^{-1}:=t(a^*_2-a)^{-1},\ \    t>0,
$$
we then conclude from \eqref{3.14} that
\begin{equation}\label{d3}
\begin{split}
\lim\limits_{a\nearrow a_2^*}\epsilon_aE_a(2)
\leq& \lim\limits_{a\nearrow a_2^*}\epsilon_a\mathcal{E}_a(\tilde{\gamma}_\tau^{(2)})
\leq\inf\limits_{t>0}\int_{\R^3} \Big(
t^2\rho_{\gamma^{(2)}}^{\frac{5}{3}}-t|x|^{-1}
\rho_{\gamma^{(2)}}\Big)dx\\
=&-\frac{1}{4}\Big(\int_{\R^3}\rho_{\gamma^{(2)}}^{\frac{5}{3}}dx\Big)^{-1}\Big(\int_{\R^3}|x|^{-1}
\rho_{\gamma^{(2)}}dx\Big)^{2}
:=-2M_1<0,
\end{split}
\end{equation}
and hence,
\begin{equation}\label{d4}
-M_1\epsilon_a^{-1}\geq E_a(2)=\mathcal{E}_a(\gamma_a)\geq \int_{\R^3}V(x)\rho_{\gamma_a}(x)dx \ \ \ \mathrm{as}\ \ a\nearrow a^*_2,
\end{equation}
where $\gamma_a=\sum_{i=1}^{2}|u_i^a\rangle\langle u_i^a|$ is a minimizer of $E_a(2)$.
Moreover, similar to \eqref{2.2},  we have
\begin{equation}\label{3.9a}
\begin{split}
E_a(2)+\int_{\R^3}V(x)\rho_{\gamma_a}dx
&=\mathcal{E}_a(\gamma_a)
+\int_{\R^3}V(x)\rho_{\gamma_a}dx\\
&\geq\ \frac{a^*_2-a}{2a^*_2}\mathrm{Tr}(-\Delta\gamma_a)- \frac{64K^2a^*_2}{a^*_2-a}\ \ \   \mathrm{as}\ \ a\nearrow a^*_2,
\end{split}
\end{equation}
where $K\in\mathbb{N}^+$ is given by \eqref{v}.
It then follows from \eqref{d4} and \eqref{3.9a} that
\begin{equation}\label{3.19}
\begin{split}
-M_1\epsilon_a^{-1}\geq&\int_{\R^3}V(x)\rho_{\gamma_a}dx
\geq E_a(2)+\int_{\R^3}V(x)\rho_{\gamma_a}dx\\
\geq&\ \frac{\epsilon_a}{2a^*_2}\mathrm{Tr}(-\Delta\gamma_a)- 64K^2a^*_2\epsilon_a^{-1}\\
\geq&\ \frac{1}{2}\epsilon_a \int_{\R^3}\rho_{\gamma_a}^{\frac{5}{3}}dx-64K^2a^*_2\epsilon_a^{-1}\\[2mm]
\geq & -64K^2a^*_2\epsilon_a^{-1}\ \ \   \mathrm{as}\,\ \ a\nearrow a^*_2.
\end{split}
\end{equation}
Together with (\ref{d3}) and \eqref{d4}, this completes the proof of Lemma \ref{lem3.1}.\qed

\vspace{.1cm}

Let  $\gamma_{a_n}=\sum_{i=1}^2|u_i^{a_n}\rangle\langle u_i^{a_n}|$ be a minimizer of $E_{a_n}(2)$, where $a_n\nearrow a^*_2$ as $n\to\infty$. Note from  \eqref{3.4} that the sequence  $\big\{\epsilon_{a_n}^2\mathrm{Tr}(-\Delta\gamma_{a_n})\big\}=\big\{\epsilon_{a_n}^2\sum_{i=1}^2\int_{\R^3}|\nabla u_i^{a_n}|^2dx\big\}$ is bounded uniformly in $n$, where $\epsilon_{a_n}:=a^*_2-a_n>0$. This yields that the sequence $\big\{\epsilon_{a_n}^{3/2}u_i^{a_n}(\epsilon_{a_n}x)\big\}_{n=1}^\infty$ is bounded uniformly in $H^1(\R^3)$, which thus admits a weak limit $w_i\in H^1(\R^3)$ for $i=1,2$.  The following lemma shows that, after suitable transformations, the operator $|w_1\rangle\langle w_1|+|w_2\rangle\langle w_2|$ is actually a minimizer of $a^*_2$.

\begin{lem}\label{lem3.2} Let  $\gamma_{a_n}=\sum_{i=1}^2|u_i^{a_n}\rangle\langle u_i^{a_n}|$ be a minimizer of $E_{a_n}(2)$, where $a_n\nearrow a^*_2$ as $n\to\infty$.
Then there exist a subsequence, still denoted by $\{u_i^{a_n}\}_{n=1}^\infty$, of $\{u_i^{a_n}\}_{n=1}^\infty$ and a point  $y_{k_*}\in \{y_1,\cdots,y_K\}$ given by \eqref{v} such that for $i=1, 2$,
\begin{equation}\label{h1}
w_i^{a_n}(x):=\epsilon_{a_n}^{3/2}u_i^{a_n}(\epsilon_{a_n}x+y_{k_*})\to w_i(x)\ \ \text{strongly\ in}\ \ H^1(\R^3)\ \  as \ \ n\to\infty,
\end{equation}
where $\epsilon_{a_n}:=a^*_2-a_n>0$, and  $\gamma:=\sum_{i=1}^{2}|w_i\rangle\langle w_i|$ is a minimizer of $a^*_2$.
\end{lem}

\noindent \textbf{Proof.}  We shall carry out the proof by three steps.

{\em Step 1.} In this step, we mainly establish the weak convergence (\ref{3.16}) of $\{u_i^{a_n}\}_{n=1}^\infty$ after transformations, where $i=1, 2$.

Note from \eqref{v} and \eqref{3.3} that
$$
\epsilon_{a_n}\sum_{k=1}^{K}\int_{\R^3}|x-y_k|^{-1}\rho_{\gamma_{a_n}}dx
= -\epsilon_{a_n}\int_{\R^3}V(x)\rho_{\gamma_{a_n}}dx\geq M_1'>0\  \ \text{ as}\ \ n\to\infty.
$$
This gives  that there exists  some point $y_{k_*}\in\{y_1,\cdots,y_K\}$  such that
\begin{equation}\label{3.5}
\epsilon_{a_n}\int_{\R^3}|x-y_{k_*}|^{-1}
\rho_{\gamma_{a_n}}dx\geq \frac{M_1'}{K}>0 \ \ \text{ as}\ \ n\to\infty.
\end{equation}
Set
\begin{eqnarray}\label{3.1M}
w_i^{a_n}(x):=\epsilon_{a_n}^{\frac{3}{2}}u_i^{a_n}(\epsilon_{a_n} x+y_{k_*}),\ \ \,  \tilde{\gamma}_{a_n}:=\sum_{i=1}^{2}|w_i^{a_n}\rangle\langle w_i^{a_n}|,\ \ \, \epsilon_{a_n}:=a^*_2-a_n>0,
\end{eqnarray}
where the point $y_{k_*}\in\{y_1,\cdots,y_K\}$ is as in  \eqref{3.5}.
We then have $(w_i^{a_n}, w_j^{a_n})=\delta_{ij}$,
\begin{eqnarray}\label{3.1}
\mathrm{Tr}(-\Delta\tilde{\gamma}_{a_n})=\sum_{i=1}^2\int_{\R^3}|\nabla w_i^{a_n}|^2dx	 =\epsilon_{a_n}^2\mathrm{Tr}(-\Delta\gamma_{a_n}),
\end{eqnarray}
and
\begin{eqnarray}\label{3.1b}
 \int_{\R^3}\rho_{\tilde{\gamma}_{a_n}}^{\frac{5}{3}}dx=\int_{\R^3}\Big(\sum_{i=1}^2|w_i^{a_n}|^2\Big)^{\frac{5}{3}}dx	 =\epsilon_{a_n}^2\int_{\R^3}\rho_{\gamma_{a_n}}^{\frac{5}{3}}dx.
\end{eqnarray}
By the uniform boundedness of the sequence $\big\{\epsilon_{a_n}^2\mathrm{Tr}(-\Delta\gamma_{a_n})\big\}$ in $n>0$, we obtain from \eqref{3.1} that
\begin{eqnarray}\label{3.22}
	\big\{w_i^{a_n}\big\}_{n=1}^\infty\ \ \text{is\  bounded\ uniformly\  in} \  H^1(\R^3),\ \ i=1,2.
\end{eqnarray}
Hence, up to a subsequence if necessary, there exists $w_i\in H^1(\R^3)$ such that
\begin{eqnarray}\label{3.16}
w_i^{a_n}\rightharpoonup  w_i\ \ \, \mathrm{  weakly \, \ in}\   \,  H^1(\R^3)\ \, \mathrm{as}\ \, n\to\infty,\ i=1,2,
\end{eqnarray}
and
\begin{eqnarray}\label{3.17}
\rho_{\tilde{\gamma}_{a_n}}\rightarrow \rho_\gamma:=w_1^2+w_2^2\ \ \, \mathrm{strongly\ \, in} \ \, L^r_{loc}(\R^3) \ \, \mathrm{as}\ \, n\to\infty,\ \, 1\leq r<3,
\end{eqnarray}
where $\gamma:=\sum_{i=1}^2|w_i\rangle\langle w_i|$.  We thus deduce from \eqref{3.5}, \eqref{3.1M} and \eqref{3.17} that, up to a subsequence if necessary,
\begin{equation}\label{3.17M}
0<\lim\limits_{n\rightarrow\infty}\epsilon_{a_n}\int_{\R^3}|x-y_{k_*}|^{-1}
\rho_{\gamma_{a_n}}dx=\lim\limits_{n\rightarrow\infty}\int_{\R^3}|x|^{-1}
\rho_{\tilde{\gamma}_{a_n}}dx=\int_{\R^3}|x|^{-1}\rho_\gamma dx,
\end{equation}
which  indicates that  $\int_{\R^3}\rho_\gamma^{5/3}dx>0$.

{\em Step 2.} This step is to prove that
\begin{eqnarray}\label{step2}
\rho_{\tilde{\gamma}_{a_n}}\rightarrow \rho_\gamma\ \   \text{strongly\,  in }\,  L^{\frac{5}{3}}(\R^3)\ \text{ as }\ n\to\infty,
\end{eqnarray}
By the Br$\acute{e}$zis-Lieb Lemma (cf. \cite{minmax}), we only need to prove  that $\lim\limits_{n\rightarrow\infty}\int_{\R^3}\rho_{\tilde{\gamma}_{a_n}}^{5/3} dx=\int_{\R^3}\rho_\gamma^{5/3} dx$. 

We first claim that
\begin{equation}\label{3:claim-0}
\{W_n\}:=\Big\{\frac{5}{3}a^*_2\rho_{\tilde{\gamma}_{a_n}}^{2/3}\Big\}\ \ \text{is\ a\ maximizing\ sequence\ of\ the best constant}\ L^*_{2},
\end{equation}
where
\begin{equation}\label{3:claim-1}
L^*_{2}:=\underset{0\leq W\in L^{5/2}(\R^3)\backslash\{0\}}{\mathrm{sup}}\frac{\sum_{j=1}^2|
\lambda_j(-\Delta-W)|}{\int_{\R^3}W^{5/2}(x)dx}
\end{equation}
is attainable (cf.\cite[Corollary 2]{iii}).
Here $\lambda_j(-\Delta-W)\leq0$ denotes the $i$th negative eigenvalue (counted with multiplicity) of $-\Delta-W(x)$ in $L^2(\R^3)$ when it exists, and zero otherwise.
Recall from  Theorem \ref{th1} that  the function $u_i^{a_n}$ satisfies
$$
H_V^{a_n}u_i^{a_n}:=\Big[-\Delta+V(x)-\frac{5}{3}a_n\Big(\sum_{j=1}^2|u_j^{a_n}|^2\Big)^{\frac{2}{3}}\Big]u_i^{a_n}
=\mu_i^{a_n}u_i^{a_n}\ \ \  \mathrm{in}\ \ \R^3,\ \   i=1,2,
$$
where $\mu_i^{a_n}<0$ is the $i$th eigenvalue (counted with multiplicity) of the operator $H_V^{a_n}$. We thus deduce from \eqref{3.1M} that $w_i^{a_n}$ solves the following system
\begin{equation}\label{wequ}
-\Delta w_i^{a_n}+\epsilon_{a_n}^2V\big(\epsilon_{a_n}x+y_{k_*}\big)w_i^{a_n}-\frac{5}{3}a_n\Big(\sum_{j=1}^2|w_j^{a_n}|^2\Big)^{\frac{2}{3}}w_i^{a_n}
=\epsilon_{a_n}^2\mu^{a_n}_i w_i^{a_n}\ \ \mathrm{in}\ \ \R^3,
\end{equation}
where $i=1,2$. Applying Lemma \ref{lem3.1}, we derive from (\ref{wequ}) that
\begin{equation}\label{3.7}
\mathrm{Tr}\big(-\Delta\tilde{\gamma}_{a_n}\big)-\frac{5}{3}a^*_2\int_{\R^3}\rho_{\tilde{\gamma}_{a_n}}^{\frac{5}{3}}dx
=\sum_{i=1}^2\epsilon_{a_n}^2\mu^{a_n}_i+o(1)\ \ \  \mathrm{as}\ \ n\rightarrow\infty,
\end{equation}
where $\tilde{\gamma}_{a_n}=\sum_{i=1}^{2}|w_i^{a_n}\rangle\langle w_i^{a_n}|$ is defined by (\ref{3.1M}).  Moreover, it follows from \eqref{3.3}, \eqref{3.1} and \eqref{3.1b}  that
\begin{equation}\label{3.6}
\begin{split}
o(1)&=\epsilon^2_{a_n}\Big(E_{a_n}(2)-\int_{\R^3}V(x)\rho_{\gamma_{a_n}}dx\Big)\\
&=\mathrm{Tr}\big(-\Delta\tilde{\gamma}_{a_n}\big)-a_n\int_{\R^3}\rho_{\tilde{\gamma}_{a_n}}^{\frac{5}{3}}dx\ \ \   \text{as}\ \ n\rightarrow\infty.
\end{split}
\end{equation}
We thus conclude from \eqref{3.7} and (\ref{3.6}) that
\begin{equation*}
\begin{split}
\liminf\limits_{n\rightarrow\infty}\Big(-\sum_{i=1}^2\epsilon_{a_n}^2\mu^{a_n}_i\Big)
=\frac{2}{3}a^*_{2}\liminf\limits_{n\rightarrow\infty}\int_{\R^3}\rho_{\tilde{\gamma}_{a_n}}^{\frac{5}{3}}dx
\geq C\int_{\R^3}\rho_\gamma^{\frac{5}{3}}dx>0,
\end{split}
\end{equation*}
where $C>0$ is independent of $n>0$.  Thus, up to a subsequence if necessary, by the min-max principle of {\cite[Sect. 12.1]{analysis}}, we deduce from \eqref{3.7} that
\begin{equation*}
\begin{split}
\sum_{j=1}^2\lambda_j\big(-\Delta-W_n\big)
&\leq \sum_{j=1}^2 \Big(\big(-\Delta-W_n\big)w_j^{a_n},\ w_j^{a_n}\Big)\\
&=\sum_{j=1}^2\epsilon_{a_n}^2\mu^{a_n}_j+o(1)\\
&=-\frac{2}{3}a^*_2\int_{\R^3}
\rho_{\tilde{\gamma}_{a_n}}^{\frac{5}{3}}dx+o(1)\ \ \   \mathrm{as}\ \ n\rightarrow\infty,
\end{split}
\end{equation*}
where $W_n=\frac{5}{3}a^*_2\rho_{\tilde{\gamma}_{a_n}}^{2/3}$ is defined by \eqref{3:claim-0}.
This further  indicates that
\begin{equation*}
\begin{split}
\frac{\sum_{j=1}^2|\lambda_j(-\Delta-W_n)|}{\int_{\R^3}W_n^{5/2}dx}
&=\ \Big(\frac{5}{3}a^*_2\Big)^{-\frac{5}{2}}
\ \frac{\sum_{j=1}^2|\lambda_j(-\Delta-W_n)|}{\int_{\R^3}\rho_{\tilde{\gamma}_{a_n}}^{5/3}dx}\\
&\geq\
\frac{2}{5}\Big(\frac{3}{5}\Big)^{\frac{3}{2}}\big(a^*_2\big)^{-\frac{3}{2}}+o(1)\\[2mm]
&=L^*_{2}+o(1)\ \ \   \mathrm{as}\ \ n\rightarrow\infty,
\end{split}
\end{equation*}
where the last identity follows from \eqref{dual}. By the definition of $L^*_{2}$ in \eqref{3:claim-1}, we then obtain that $\{W_n\}=\big\{\frac{5}{3}a^*_2\rho_{\tilde{\gamma}_{a_n}}^{2/3}\big\}$ is a maximizing sequence of $L^*_{2}$, and the claim (\ref{3:claim-0}) is thus established.

We now denote
$$
\alpha:=\lim\limits_{n\rightarrow\infty}\int_{\R^3}\rho_{\tilde{\gamma}_{a_n}}^{5/3} dx\ \ \ \text{and}  \ \ \ \beta:=\int_{\R^3}\rho_\gamma^{5/3} dx,
$$
where $\alpha\geq\beta>0$ holds true in view of (\ref{3.17M}).
To establish Step 2, on the contrary, suppose that $\alpha>\beta>0$. By an adaptation of the classical dichotomy result (cf. \cite[Sect. 3.3]{begain}),  there exist a subsequence, still denoted by $\{\rho_{\tilde{\gamma}_{a_n}}\}$,  of $\{\rho_{\tilde{\gamma}_{a_n}}\}$ and a sequence $\{R_n\}$ with $R_n\rightarrow\infty$ as $n\to\infty$ such that
\begin{equation*}
\label{17}0<\lim\limits_{n\to\infty}\int_{|x|\leq R_n}\rho_{\tilde{\gamma}_{a_n}}^{\frac{5}{3}}dx=\beta,\ \ \ \ \  \lim\limits_{n\to\infty}\int_{R_n\leq|x|\leq 2R_n}\rho_{\tilde{\gamma}_{a_n}}^{\frac{5}{3}}dx=0.
\end{equation*}
Thus, the argument of \cite[Lemma 17]{iii} yields that
there exists some $s\in \{0,\ 1,\ 2\}$ such that
\begin{equation}\label{3.31a}
\begin{split}
\sum_{j=1}^2\big|\lambda_j(-\Delta-W_n)\big|
=&\sum_{j=1}^s\big|\lambda_j\big(-\Delta-W_n\mathbf{1}_{B_{R_n}}\big)\big|\\
&+\sum_{j=1}^{2-s}\big|\lambda_j\big(-\Delta-W_n\mathbf{1}_{\R^3\backslash B_{2R_n}}\big)\big|+o(1)\ \ \ \mathrm{as}\ \ n\rightarrow\infty,
\end{split}
\end{equation}
where $W_n=\frac{5}{3}a^*_2\rho_{\tilde{\gamma}_{a_n}}^{2/3}$ for all $n>0$.
Recall from \eqref{lt} that  the best constant $L^*_s$ of the  finite rank Lieb-Thirring inequality is defined as
\begin{equation*}
L^*_s:=\underset{0\leq W\in L^{5/2}(\R^3)\backslash\{0\}}{\mathrm{sup}}\frac{\sum_{j=1}^s|
\lambda_j(-\Delta-W)|}{\int_{\R^3}W^{5/2}(x)dx},\ \   \forall \ s\in\mathbb{N}.
\end{equation*}
According to the above definition, it is obvious that $L^*_{s}$ is increasing in $s>0$. Together with \eqref{3.31a}, one hence gets from the claim \eqref{3:claim-0} that for some $s\in \{0,\ 1,\ 2\}$,
\begin{align}\label{3.27}
\ \Big(\frac{5}{3}a^*_2\Big)^{\frac{5}{2}}\alpha L^*_{2}
=&\  L^*_{2}\lim\limits_{n\rightarrow\infty}\int_{\R^3}W_n^{\frac{5}{2}}(x)dx\nonumber\\
=&\lim\limits_{n\rightarrow\infty}\sum_{j=1}^2\big|\lambda_j\big(-\Delta-W_n\big)\big|\\[1mm]
\leq&\ L^*_{s}\lim\limits_{n\rightarrow\infty}\int_{\R^3}\big(W_n\mathbf{1}_{B_{R_n}}\big)^{\frac{5}{2}}dx
+L^*_{2-s}\lim\limits_{n\rightarrow\infty}\int_{\R^3}\big(W_n\mathbf{1}_{\R^3\backslash B_{2R_n}}\big)^{\frac{5}{2}}dx\nonumber\\[2mm]
=&\ \Big(\frac{5}{3}a^*_2\Big)^{\frac{5}{2}}
\Big[L^*_{s}\beta+L^*_{2-s}(\alpha-\beta)\Big]\leq \Big(\frac{5}{3}a^*_2\Big)^{\frac{5}{2}}\alpha L^*_{2},\nonumber
\end{align}
where  the last inequality follows from the fact that $L^*_{2-s}\leq L^*_{2}$ and $L^*_{s}\leq L^*_{2}$ hold for $s\in\{0,\,1,\,2\}$.  Since $\alpha>\beta>0$, we  obtain  from \eqref{3.27} that $L^*_{s}=L^*_{2-s}=L^*_{2}$, where  $s\in \{0,\, 1,\, 2\}$ is as in \eqref{3.27}. However, recalling from \eqref{strict} (or see \cite[Theorem 6]{ii}) that $a_1^*>a_2^*>0$, one can conclude from \eqref{dual} that $0<L^*_{1}<L^*_{2}$, which gives that $s\neq1$. Moreover,   because $L^*_{0}=0$, it further yields that  $s\neq0,2$. These thus lead to a contradiction, if $\alpha>\beta>0$. This implies that $\lim\limits_{n\rightarrow\infty}\int_{\R^3}\rho_{\tilde{\gamma}_{a_n}}^{5/3} dx=\int_{\R^3}\rho_\gamma ^{5/3} dx$, and Step 2 is therefore established.   

{\em Step 3.} Since Step 2 gives that  $\rho_{\tilde{\gamma}_{a_n}}\rightarrow \rho_\gamma$ strongly  in $L^{\frac{5}{3}}(\R^3)$ as $n\to\infty$, we derive from \eqref{3.16} and \eqref{3.6}  that
\begin{align}\label{3.31}
a^*_2\int_{\R^3}\rho_\gamma^{\frac{5}{3}}dx
=&\ a^*_2\lim\limits_{n\rightarrow\infty}\int_{\R^3}\rho_{\tilde{\gamma}_{a_n}}^{\frac{5}{3}}dx
=\liminf\limits_{n\rightarrow\infty}\mathrm{Tr}(-\Delta\tilde{\gamma}_{a_n})\nonumber\\
\geq&\ \mathrm{Tr}(-\Delta\gamma)\geq a^*_2\|\gamma\|^{-\frac{2}{3}}\int_{\R^3}\rho_\gamma^{\frac{5}{3}}dx\\
\geq& \ a^*_2\int_{\R^3}\rho_\gamma^{\frac{5}{3}}dx,\nonumber
\end{align}
where we have used the definition of $a^*_2$, together with the fact that $\|\gamma\|\leq\liminf\limits_{n\rightarrow\infty}\|\tilde{\gamma}_{a_n}\|=1$. This further shows that $\|\gamma\|=1$, $\gamma$ is an optimizer of $a^*_2$, and
\begin{eqnarray}\label{3.33}
|\nabla w_i^{a_n}|\rightarrow |\nabla w_i|\ \ \, \text{strongly\  in} \ \, L^2(\R^3)\ \, \text{as} \ \, n\to\infty,\ \,   i=1,2.
\end{eqnarray}

Since   $\gamma$ is an optimizer of $a^*_2$, using again the fact (cf. \cite[Proposition 11]{ii}) that  $0<a^*_2<a^*_1$, we obtain that Rank$(\!\gamma) =2$. This further implies from \eqref{form} that
\begin{eqnarray}\label{3.34a}
\int_{\R^3}\rho_{\gamma}dx=\|\gamma\|\, \text{Rank}(\!\gamma) =2.
\end{eqnarray}
We thus deduce from \eqref{3.16} and \eqref{3.34a} that
\begin{eqnarray}\label{3.34}
\rho_{\tilde{\gamma}_{a_n}}=\sum_{i=1}^2|w_i^{a_n}|^2\rightarrow \rho_\gamma=\sum_{i=1}^2w_i^2\ \ \, \text{strongly \ in} \ \, L^{1}(\R^3)\ \text{ as }\ n\to\infty,
\end{eqnarray}
where $\gamma=\sum_{i=1}^2|w_i\rangle\langle w_i|$ is a minimizer of $a^*_2$.
Using again the Br$\acute{e}$zis-Lieb Lemma (cf. \cite{minmax}), one can deduce from \eqref{3.16} and \eqref{3.34} that
\begin{eqnarray}\label{3.35}
w_i^{a_n}\rightarrow  w_i\ \ \ \text{strongly\  in} \ \ L^2(\R^3)\ \ \text{as} \ \ n\to\infty,\ \,  i=1,2.
\end{eqnarray}
Together with \eqref{3.33}, this proves the $H^1-$convergence of \eqref{h1},  and we are therefore done.\qed

\subsection{Proof of Theorem \ref{th3}}

The main purpose of this subsection is to complete the proof of Theorem \ref{th3}. We first establish  the following uniformly exponential decay of the sequence $\{w_i^{a_n}\}_{n=1}^\infty$ as $n\to\infty$ for $i=1,2$.

\begin{lem}\label{lem3.3}
Suppose that the system $\big(w_1^{a_n}, w_2^{a_n}\big)$ is given by Lemma \ref{lem3.2}. Then  there exist constants $\theta>0$ and $C(\theta)>0$, which are independent of $n>0$, such that for sufficiently large $n>0$,
\begin{equation}\label{3.04}
|w_i^{a_n}(x)|\leq C(\theta)e^{-\theta|x|}\ \ \ \text{ uniformly\ in}\ \ \R^3,\ \ i=1, 2.
\end{equation}
\end{lem}

\noindent{\bf Proof.}
By the uniform boundedness \eqref{3.22} of $\{w_i^{a_n}\}_{n=1}^\infty$ in $H^1(\R^3)$ for $i=1,2$, and  the strong convergence \eqref{step2} of  $\{\rho_{\tilde{\gamma}_{a_n}}\}=\big\{\sum_{j=1}^2|w_j^{a_n}|^2\big\}$  in $L^{\frac{5}{3}}(\R^3)$, we first claim that, up to a subsequence if necessary,
\begin{equation}\label{3.23}
\sup\limits_{n>0}\|\rho_{\tilde{\gamma}_{a_n}}\|_{\infty}:=\sup\limits_{n>0}\big\|\sum_{i=1}^2|w_i^{a_n}|^2\big\|_{\infty}<+\infty,
\end{equation}
and
\begin{equation}\label{3.24}
\lim\limits_{|x|\to\infty}\rho_{\tilde{\gamma}_{a_n}}(x)=\lim\limits_{|x|\to\infty}\sum_{i=1}^2|w_i^{a_n}|^2=0 \ \
\ \mathrm{uniformly\ for \ sufficiently\ large}\ n>0,
\end{equation}
where $\tilde{\gamma}_{a_n}=\sum_{i=1}^{2}|w_i^{a_n}\rangle\langle w_i^{a_n}|$.

Actually, note  from \eqref{wequ} that the function $w_i^{a_n}$ satisfies
\begin{equation}\label{3.18}
\big(-\Delta -c_{a_n}(x)\big)w_i^{a_n}
=\epsilon_{a_n}^2\mu^{a_n}_i w_i^{a_n}\ \ \  \mathrm{in}\ \ \R^3, \ \ i=1,\, 2,
\end{equation}
where
$$
c_{a_n}(x)=\epsilon_{a_n}\sum_{k=1}^K\big|x-\epsilon^{-1}_{a_n} (y_k-y_{k_*})\big|^{-1}+\frac{5}{3}a_n\Big(\sum_{j=1}^2|w_j^{a_n}|^2\Big)^{2/3},
$$
and $\mu^{a_n}_1<\mu^{a_n}_2<0$ holds for all $n>0$.
We then obtain from Kato's  inequality (cf. \cite[Theorem X.27]{modern}) that
\begin{equation}\label{hq}
\big(-\Delta-c_{a_n}(x)\big)|w^{a_n}_i|\leq0\ \ \ \ \mathrm{in}\ \ \R^3,\ \ i=1,2.
\end{equation}
Following the uniform boundedness \eqref{3.22} of $\{w_i^{a_n}\}_{n=1}^\infty$  in $H^1(\R^3)$ for $i=1, 2$, one can verify that
$$
\|c_{a_n}(x)\|_ {L^r(B_2(y))}\leq C\ \ \, \mathrm{holds\  for\  any}\ \,  y\in\R^3,
$$
where  $r\in\big(3/2, 3\big)$, and $C>0$ is independent of $n>0$ and $y\in\R^3$. Thus, applying De Giorgi-Nash-Moser theory (cf. \cite[Theorem 4.1]{hq}),  we immediately conclude from \eqref{hq}  that
\begin{equation}\label{3.41}
\begin{split}
\|w_i^{a_n}\|_{L^\infty(B_1(y))}&\leq C_1\|w_i^{a_n}\|_{L^{10/3}(B_2(y))}\\[1.5mm]
&\leq C_1\|\rho_{\tilde{\gamma}_{a_n}}\|^{1/2}_{L^{5/3}(B_2(y))}\ \ \ \ \mathrm{for\ \, any}\ \,   y\in\R^3,\ \ \ i=1,2,
\end{split}
\end{equation}
where $\rho_{\tilde{\gamma}_{a_n}}=\sum_{i=1}^2|w_i^{a_n}|^2$, and $C_1>0$ is  independent of $n>0$ and $y\in\R^3$. By  the strong convergence \eqref{step2} of  $\rho_{\tilde{\gamma}_{a_n}}$ in $L^{\frac{5}{3}}(\R^3)$,  we thus obtain  from \eqref{3.41} that  both \eqref{3.23} and \eqref{3.24} hold true, and the above  claim is therefore proved.

Furthermore, it follows from \eqref{3.18} that
\begin{equation}\label{mu}
\begin{split}
\sum_{i=1}^{2}\epsilon_{a_n}^2\mu_i^{a_n}
=&\ \mathrm{Tr}(-\Delta\tilde{\gamma}_{a_n})-\frac{5}{3}a_n\int_{\R^3}\rho_{\tilde{\gamma}_{a_n}}^{\frac{5}{3}}dx\\
&\ -\epsilon_{a_n}\sum_{k=1}^K\int_{\R^3}\big|x-\epsilon^{-1}_{a_n} (y_k-y_{k_*})\big|^{-1}\rho_{\tilde{\gamma}_{a_n}}dx,
\end{split}
\end{equation}
and
\begin{equation}\label{3.20}
w_i^{a_n}(x)=\int_{\R^3}G_i^{a_n}(x-y)\Big[\epsilon_{a_n}\sum_{k=1}^{K}\big|y-\epsilon_{a_n}^{-1} (y_k-y_{k_*})\big|^{-1}+\frac{5}{3}a_n\rho_{\tilde{\gamma}_{a_n}}^{\frac{2}{3}}(y)\Big]w_i^{a_n}(y)dy,
\end{equation}
where $\rho_{\tilde{\gamma}_{a_n}}=\sum_{j=1}^2|w_j^{a_n}|^2$,  and $G_i^{a_n}(x)$ denotes the Green's function of the operator  $$-\Delta-\epsilon_{a_n}^2\mu_i^{a_n}\ \ \text{ in}\ \  \R^3.$$  
Using again the uniform boundedness  of $\{w_i^{a_n}\}_{n=1}^\infty$  in $H^1(\R^3)$ for $i=1,2$, we derive  from \eqref{mu} that the sequence  $\big\{\sum_{i=1}^{2}\epsilon_{a_n}^2\mu_i^{a_n}\big\}$ is  bounded uniformly in $n>0$. Since $\mu_1^{a_n}<\mu_2^{a_n}<0$, this implies that
\begin{align}\label{3.49a}
\big\{\epsilon_{a_n}^2\mu_i^{a_n}\big\}_{n=1}^\infty\ \text{is\ also\ bounded\ uniformly\ in} \ n>0,\ \ i=1, 2.
\end{align}
Thus,  up to a subsequence if necessary, we can assume that
$$\lim\limits_{n\rightarrow\infty}\epsilon_{a_n}^2\mu_i^{a_n}=\hat{\mu}_i\leq0.$$
Passing to the limit on both hand sides of \eqref{3.18} as $n\to\infty$, we then obtain from Lemma \ref{lem3.2} that
\begin{align}\label{3.46}
-\Delta w_i-\frac{5}{3}a^*_2\Big(\sum_{j=1}^2w_j^2\Big)^{\frac{2}{3}}w_i=\hat{\mu}_i w_i\ \ \, \mathrm{in}\ \, \R^3,\ \, i=1,2,
\end{align}
where $w_i$ is the strong limit of $w_i^{a_n}$ in $H^1(\R^3)$. Recall from Lemma \ref{lem3.2} that  the  functions $w_1$ and $w_2$  satisfy $(w_i,w_j)=\delta_{ij}$, and $\gamma:=\sum_{i=1}^2|w_i\rangle \langle w_i|$ is a minimizer of $a^*_2$.
We then conclude from \eqref{nls} (or \cite[ Theorem 6]{ii}) that $\hat{\mu}_i<0$ holds for $i=1,2$.
As a consequence, employing  the fact (cf. \cite[Theorem 6.23]{analysis}) that
$$
G_i^{a_n}(x)=\frac{1}{4\pi|x|}e^{-\sqrt{|\epsilon_{a_n}^2\mu_i^{a_n}|}|x|}\ \ \ \ \mathrm{in}\ \ \R^3,
$$
we deduce from \eqref{3.20} that for any sufficiently large $n>0$, 
\begin{align}\label{3.21}
|w_i^{a_n}(x)|\leq&\ C\int_{\R^3}
|x-y|^{-1}
e^{{-\theta}|x-y|}|w_i^{a_n}(y)|\\
&\ \ \ \ \cdot\Big(\epsilon_{a_n}\sum_{k=1}^{K}\big|y-\epsilon_{a_n}^{-1} (y_k-y_{k_*})\big|^{-1}+\frac{5}{3}a^*_{2}\rho_{\tilde{\gamma}_{a_n}}^{\frac{2}{3}}(y)\Big)dy\ \ \mathrm{in}\  \R^3,\ i=1,2,\nonumber
\end{align}
where $\theta:=\frac{1}{2}\min\big\{\sqrt{|\hat{\mu}_1|},\sqrt{|\hat{\mu}_2|}\big\}>0$, and  $C>0$ is independent of $n>0$.

Following  \eqref{3.23},  \eqref{3.24} and \eqref{3.21},  the exponential decay of \eqref{3.04} can be proved in a similar way of {\cite[Lemma 3.3]{me}}, and we omit the detailed proof for simplicity.  This ends the proof of Lemma \ref{lem3.3}. \qed
\vspace{.1cm}

\noindent\textbf{Proof  of Theorem \ref{th3}.} Let $\gamma_{a_n}=\sum_{i=1}^2|u_i^{a_n}\rangle\langle u_i^{a_n}|$ be a minimizer of $E_{a_n}(2)$, and suppose
\begin{equation*}\label{3.48}
\tilde{\gamma}_{a_n}:=\sum_{i=1}^{2}|w_i^{a_n}\rangle\langle w_i^{a_n}|:=\sum_{i=1}^{2}\epsilon_{a_n}^{3}\big|u_i^{a_n}(\epsilon_{a_n}\cdot+y_{k_*})\big\rangle\big\langle u_i^{a_n}(\epsilon_{a_n}\cdot+y_{k_*})\big|
\end{equation*}
is as in Lemma \ref{lem3.2}, where  $\epsilon_{a_n}=a_2^*-a_n>0$ and $a_n\nearrow a_2^*$ as $n\to\infty$. The $H^1$-uniform convergence of \eqref{th1.3.1} then follows directly from Lemma \ref{lem3.2}.

We now prove the energy estimate \eqref{th1.3.2}. Indeed, by the definition of $a^*_2$,  it follows from \eqref{step2} that
\begin{equation}\label{3.36a}
\begin{split}
\epsilon_{a_n}E_{a_n}(2)
=&\epsilon_{a_n}\mathcal{E}_{a_n}(\gamma_{a_n})\\
\geq& \epsilon_{a_n}\big(a^*_2-a\big)\int_{\R^3}\rho^{\frac{5}{3}}_{\gamma_{a_n}}dx-\epsilon_{a_n}\int_{\R^3}\sum_{k=1}^K|x-y_k|^{-1}\rho_{\gamma_{a_n}}dx\\
 =&\int_{\R^3}\rho_{\tilde{\gamma}_{a_n}}^{\frac{5}{3}}dx-\int_{\R^3}\sum_{k=1}^K\big|x+\epsilon_{a_n}^{-1}(y_{k_*}-y_k)\big|^{-1}\rho_{\tilde{\gamma}_{a_n}}dx\\
=&\int_{\R^3}\rho^{\frac{5}{3}}_{\gamma}dx-\int_{\R^3} |x|^{-1}\rho_{\gamma}dx+o(1)\ \ \ \mathrm{as}\ \ n\rightarrow\infty,
\end{split}
\end{equation}
where $y_{k_*}\in\{y_1,\cdots,y_K\}$ and $\gamma=\sum_{i=1}^{2}|w_i\rangle\langle w_i|$  are as in Lemma \ref{lem3.2}.
Since $\gamma=\sum_{i=1}^{2}|w_i\rangle\langle w_i|$ satisfying $(w_i,w_j)=\delta_{ij}$ is a minimizer of $a_2^*$, we obtain from \eqref{d3} and   \eqref{3.36a} that
\begin{align}\label{3.21a}
\int_{\R^3}\Big(\rho_{\gamma}^{\frac{5}{3}}-|x|^{-1}\rho_{\gamma}\Big)dx=\lim\limits_{n\to\infty}\epsilon_{a_n}E_{a_n}(2)=\inf\limits_{t>0}\int_{\R^3}\Big(t^2\rho_{\gamma}^{\frac{5}{3}}-t|x|^{-1}\rho_{\gamma}\Big)dx.
\end{align}
Note that the right-hand side of \eqref{3.21a} has exactly one optimizer
$$
t_{min}=\Big(2\int_{\R^3}\rho^{5/3}_{\gamma}dx\Big)^{-1}\int_{\R^3}|x|^{-1}\rho_{\gamma}dx.
$$
We thus conclude from \eqref{3.21a} that
\begin{equation}\label{3.37}
1=\Big(2\int_{\R^3}\rho^{5/3}_{\gamma}dx\Big)^{-1}\int_{\R^3}|x|^{-1}\rho_{\gamma}dx,
\end{equation}
which further yields in turn  that
\begin{equation}\label{3.38}
\lim\limits_{n\to\infty}\epsilon_{a_n}E_{a_n}(2)=-\frac{1}{2}\int_{\R^3}|x|^{-1}\rho_{\gamma}dx=-\int_{\R^3}\rho^{5/3}_{\gamma}dx=-\frac{1}{a_2^*}\mathrm{Tr}\big(-\Delta \gamma\big).
\end{equation}
Here we have used  $\|\gamma\|=1$ and the fact that $\gamma$ is a minimizer of $a_2^*$. This proves  \eqref{th1.3.2}.


We next prove the $L^\infty$-uniform convergence \eqref{th1.3.1} of $w_i^{a_n}$ as $n\to\infty$, i.e.,
\begin{equation}\label{3.51}
w_i^{a_n}\to w_i\ \ \ \ \text{strongly\ in}\ \ L^\infty(\R^3)\ \ \mathrm{as}\ \ n\rightarrow\infty, \ \ i=1,2,
\end{equation}
where  the system $(w^{a_n}_1, w^{a_n}_2)$ is defined by Lemma \ref{lem3.2}.
Note from \eqref{3.18} that $w_i^{a_n}$ satisfies
\begin{align}\label{3.53}
-\Delta w_i^{a_n}=&\epsilon_{a_n}\sum_{k=1}^K\big|x-\epsilon^{-1}_{a_n} (y_k-y_{k_*})\big|^{-1}w_i^{a_n}\nonumber\\
&
+\frac{5}{3}a_n\Big(\sum_{j=1}^2|w_j^{a_n}|^2\Big)^{\frac{2}{3}}w_i^{a_n}+\epsilon_{a_n}^2\mu_i^{a_n} w_i^{a_n}\\
:=&f_i^n(x)\ \ \ \   \mathrm{in}\ \ \R^3,\ \ i=1,2,\nonumber
\end{align}
and the sequence $\{f_i^n(x)\}_{n=1}^\infty$ is  bounded uniformly in $L^{2}_{loc}(\R^3)$ for  $i=1,2$ in view of \eqref{3.22}, \eqref{3.23} and  \eqref{3.49a}. One hence gets  from \eqref{3.53} and \cite[Theorem 8.8]{elli}  that for any fixed $ R>0$,
$$
\|w_i^{a_n}\|_{W^{2, 2}(B_R)}\leq C\Big(\|w_i^{a_n}\|_{H^{1}(B_{R+1})}
+ \|f_i^n\|_{L^{2}(B_{R+1})}   \Big),\ \  \ i=1,2,
$$
where
$C>0$ is independent of $n>0$. This implies that $\{w_i^{a_n}\}_{n=1}^\infty$ is bounded uniformly  in $W^{2, 2}(B_R)$ for  $i=1,2$. Consequently, by the compact embedding theorem (cf.\cite[Theorem 7.26]{elli}) from  $W^{2,2}(B_R)$ into $ L^\infty(B_R)$, we obtain  that there exists a subsequence, still denoted by $\{w_i^{a_n}\}_{n=1}^\infty$, of $\{w_i^{a_n}\}_{n=1}^\infty$ such that for any  fixed $R>0$,
\begin{equation}\label{conv1}
w_i^{a_n}\rightarrow w_i\ \ \,   \mbox{strongly in}\ \, L^\infty(B_R)\ \ \mathrm{as}\ n\to\infty,\ \ i=1,2.
\end{equation}

On the other hand,  since $\sum_{i=1}^{2}|w_i\rangle\langle w_i|$ is a minimizer of $a_2^*$,   we get from \eqref{3.24} and the exponential decay   \eqref{3.0} that for any $\varepsilon>0$,  there exists a sufficiently large constant $R:=R(\varepsilon) > 0$, which is independent of $n>0$, such that for sufficiently large $n>0$,
\begin{eqnarray}\label{3.32}
|w_i(x)|,\ \    |w_i^{a_n}(x)|<\frac{\varepsilon}{4}\ \   \mathrm{in}\ \, \R^3\backslash B_R,\ \ i=1,2,
\end{eqnarray}
and hence,
\begin{eqnarray}\label{3.56a}
\sup\limits_{|x|\geq R}\big|w_i^{a_n}(x)-w_i(x)\big|\leq	\sup\limits_{|x|\geq R}\big(|w_i^{a_n}(x)|+|w_i(x)|\big)<\frac{\varepsilon}{2},\ \ i=1,2.
\end{eqnarray}
Together with \eqref{conv1}, we obtain from \eqref{3.56a} that the convergence of \eqref{3.51} is  true, and the proof of Theorem \ref{th3} is therefore complete. \qed

\begin{rem}\label{rem3.1}
Generally, suppose there exists an integer $2\leq N\in\mathbb{N}^+$ such that $a^*_{N-1}>a^*_{N}$ holds, which is true at least for $N=2$. It then follows from \eqref{form} that any minimizer $\gamma^{(N)}$ of $a_N^*$  can be written in the form $\gamma^{(N)}=\|\gamma^{(N)}\|\sum_{i=1}^N|Q_i\rangle\langle Q_i|$, where $(Q_i,Q_j)=\delta_{ij},\ i,j=1,\cdots,N$. Since $a_k^*>a_{2k}^*$ holds for any $k\in\mathbb{N}^+$  (see \cite[Proposition 11]{ii}),  the same argument of proving Theorem \ref{th3} yields that Theorem \ref{th3} essentially holds true for  any $E_a(N)$, as soon as $2\leq N\in\mathbb{N}^+$ satisfies $a^*_{N-1}>a^*_{N}$.
\end{rem}

\section{$N=3$: Limiting Behavior of Minimizers as $a\nearrow a^*_3$}
In this section, we prove Theorem \ref{th4} on the limiting behavior of minimizers  for $E_a(3)$ as $a\nearrow a^*_3$. The main idea of the proof is called the blow-up analysis of many-body fermionic problems, which is explained briefly in Subsection 1.1.

\vskip 0.05truein

\noindent\textbf{Proof of  Theorem \ref{th4}.} One can note from \eqref{k} that $a_2^*\geq a_3^*$. If $a^*_2>a^*_3$, then it follows from Remark \ref{rem3.1} that
\begin{equation}\label{M:3.36}
\mbox{Theorem \ref{th3}  holds true for $E_a(3)$ in this case. }
\end{equation}
Thus, in the following it suffices to focus on the case where $a^*_2=a^*_3$.

Let $\gamma_{a_n}=\sum_{i=1}^{3}|u_i^{a_n}\rangle\langle u_i^{a_n}|
$ be a minimizer of $E_{a_n}(3)$ with $a_n\nearrow a^*_3$ as $n\to\infty$, where we assume $a^*_2=a^*_3$. We first address the $L^\infty$-uniform convergence of $u_i^{a_n}$ as $n\to\infty$ after suitable transformations. Recall from  \eqref{d3} that
\begin{equation}\label{3.36}
\lim\limits_{a\nearrow a_2^*}(a_2^*-a)E_a(2)\leq\inf\limits_{t>0}\int_{\R^3} \Big(t^2\rho_{\gamma^{(2)}}^{\frac{5}{3}}-t|x|^{-1}
\rho_{\gamma^{(2)}}\Big)dx:=-2M_1<0,
\end{equation}
where $M_1>0$ is independent of $a\in(0,a_2^*)$, and $\gamma^{(2)}$ is a minimizer of $a_2^*$.
Since we consider the case where  $a_2^*=a_3^*$, by Lemma \ref{lem2.1} (2), we obtain from \eqref{k} and \eqref{3.36}  that
\begin{equation}\label{3.63}
\begin{split}
-M_1\epsilon_{a_n}^{-1}\geq E_{a_n}(2)\geq E_{a_n}(3)=\mathcal{E}_{a_n}(\gamma_{a_n})\geq\int_{\R^3}V(x)\rho_{\gamma_{a_n}}dx\ \ \ \mathrm{as}\ \ n\to\infty,
\end{split}
\end{equation}
where $\epsilon_{a_n}:=a_3^*-{a_n}=a_2^*-{a_n}>0$ and $\rho_{\gamma_{a_n}}=\sum_{i=1}^3|u_i^{a_n}|^2$.  Thus, similar to \eqref{3.9a} and \eqref{3.19}, we can deduce from \eqref{3.63} that the estimates of Lemma \ref{lem3.1} are also applicable to  $\gamma_{a_n}$ as $n\to\infty$, and thus we particularly have
\begin{gather}\label{3.63a}
 0<M_1\leq -\epsilon_{a_n}\int_{\R^3}V(x)\rho_{\gamma_{a_n}}dx\leq M_2\ \ \ \mbox{as}\ \ n\to\infty,
\end{gather}
where $0<M_1<M_2$ are independent of $n>0$.

Therefore,  using again the fact (cf.\cite[Proposition 11]{ii}) that $a_k^*>a_{2k}^*$ holds for any $k\in\mathbb{N}^+$, the same arguments of proving  \eqref{step2} and \eqref{3.33}  yield that, up to a subsequence if necessary,  there exist a point $y_{k_*}\in\{y_1,\cdots,y_K\}$  and $w_i\in H^1(\R^3)$ such that for $\epsilon_{a_n}:=a_3^*-{a_n}>0$,
\begin{eqnarray}\label{3.42}
w_i^{a_n}:=\epsilon_{a_n}^{\frac{3}{2}}u_i^{a_n}(\epsilon_{a_n} x+y_{k_*})\rightharpoonup  w_i\ \ \mathrm{  weakly \ in }\ H^1(\R^3)\ \ \mathrm{as}\ \ n\to\infty,\ i=1,2,3,
\end{eqnarray}
\begin{eqnarray}\label{3.45}
\rho_{\tilde{\gamma}_{a_n}}=\sum_{i=1}^3|w_i^{a_n}|^2\rightarrow \rho_\gamma=\sum_{i=1}^3w_i^2\ \ \ \text{ strongly\  in }\  L^{\frac{5}{3}}(\R^3)\ \text{ as}\ n\to\infty,
\end{eqnarray}
and
\begin{eqnarray}\label{3.44}
|\nabla w_i^{a_n}|\rightarrow |\nabla w_i|\ \ \ \text{strongly\  in} \ \ L^2(\R^3)\ \ \text{as} \ \ n\to\infty,\ \ \  i=1,2,3,
\end{eqnarray}
where $\tilde{\gamma}_{a_n}:=\sum_{i=1}^3|w_i^{a_n}\rangle\langle w_i^{a_n}|$, and
\begin{eqnarray}\label{3.62}
\gamma:=\sum_{i=1}^3|w_i\rangle\langle w_i|\ \text{ satisfying} \ \|\gamma\|=1\ \text{is\ an\ optimizer\ of}\ a^*_3.
\end{eqnarray}
Similar to \eqref{3.51}, applying the uniform boundedness of $\big\{w_i^{a_n}\big\}_{n=1}^\infty $ in $H^1(\R^3)$ for $i=1,2,3$, thus  one can further derive from \eqref{3.45} that
\begin{eqnarray}\label{3.64}
w_i^{a_n}\to  w_i\ \ \  \ \mathrm{  strongly \ in}\ \ L^\infty(\R^3)\ \ \mathrm{as}\ \ n\to\infty,\ \ i=1,2,3.
\end{eqnarray}
This proves \eqref{th4.1}.

We next analyze the properties of the limiting function $(w_1, w_2,w_3)$. Since $a_1^*>a_2^*=a_3^*$ and $\gamma=\sum_{i=1}^3|w_i\rangle\langle w_i|$ is an optimizer of $ a^*_3$,  we conclude from \eqref{k} that either Rank$(\!\gamma) =3$ or Rank$(\!\gamma) =2$. We shall discuss separately the following two different situations:


1. $a_2^*=a_3^*$ and Rank$(\!\gamma) =3$. In this situation, we  deduce from \eqref{form} and \eqref{3.62} that
\begin{eqnarray}\label{3.43}
\begin{split}
\int_{\R^3}\rho_{\tilde{\gamma}_{a_n}}dx&=\sum_{i=1}^3\int_{\R^3}|w_i^{a_n}|^2dx=\sum_{i=1}^3\int_{\R^3}|u_i^{a_n}|^2dx\\
&=3 =\|\gamma\|\, \text{Rank}(\!\gamma) =\int_{\R^3}\rho_\gamma dx=\sum_{i=1}^3\int_{\R^3}w_i^2dx.
\end{split}
\end{eqnarray}
We thus obtain from \eqref{3.42}, \eqref{3.44} and  \eqref{3.43} that
\begin{eqnarray}\label{3.49}	
w_i^{a_n}\to  w_i\ \ \ \  \mathrm{  strongly \ in }\ \ H^1(\R^3)\ \ \mathrm{as}\ \ n\to\infty,\ \ i=1,2,3,
\end{eqnarray}
which then implies from \eqref{1.2} that  the functions $w_1,w_2$ and $w_3$ satisfy $(w_i, w_j)=\delta_{ij}$.
Together with (\ref{M:3.36}), this therefore proves  Theorem \ref{th4} (1).

2. $a_2^*=a_3^*$ and  Rank$(\!\gamma) =2$. In this situation, recall from Theorem \ref{th1} and \eqref{3.42} that $u_i^{a_n}$ satisfies
\begin{equation}\label{3.64a}
H_V^{a_n}u_i^{a_n}:=\Big[-\Delta +V(x)- \frac{5}{3}a_n\Big(\sum_{j=1}^3|u_j^{a_n}|^2\Big)^{\frac{2}{3}}\Big]u_i^{a_n}=\mu_i^{a_n}u_i^{a_n}\ \ \ \text{in}\ \ \R^3,\ \ i=1,2,3,
\end{equation}
and hence for $i=1,2,3$,
\begin{equation}\label{3.72}
-\Delta w_i^{a_n}+\epsilon_{a_n}^2V\big(\epsilon_{a_n}x+y_{k_*}\big)w_i^{a_n}-\frac{5}{3}a_n\Big(\sum_{j=1}^3|w_j^{a_n}|^2\Big)^{\frac{2}{3}}w_i^{a_n}
=\epsilon_{a_n}^2\mu^{a_n}_i w_i^{a_n}\ \ \mathrm{in}\ \ \R^3,
\end{equation}
where $\mu_1^{a_n}<\mu_2^{a_n}\leq\mu_3^{a_n}<0$ are the $3$-first eigenvalues (counted with multiplicity) of $H_V^{a_n}$ in $\R^3$.

Similar to \eqref{3.46}, we then deduce from  \eqref{3.42} and \eqref{3.72} that
\begin{equation}\label{3.55}
\hat H_{\gamma} w_i:=\Big[-\Delta -\frac{5}{3}a^*_3\Big(\sum_{j=1}^3w_j^2\Big)^{\frac{2}{3}}\Big]w_i=\hat{\mu}_i w_i\ \ \   \mathrm{in}\ \, \R^3,\ \ i=1,2,3,
\end{equation}
where $\hat{\mu}_i$ satisfies, up to a subsequence if necessary,
\begin{equation}\label{3.73}
\hat{\mu}_i=\lim\limits_{n\to\infty}\epsilon_{a_n}^2\mu_i^{a_n}\ \ \text{for}\ \, i=1,2,3, \ \   \text{and}  \ \,   \hat{\mu}_1\leq\hat{\mu}_2\leq\hat{\mu}_3\leq0.
\end{equation}
As a consequence of \eqref{3.55}, we  conclude  that for $i\in\{1,2,3\}$,
\begin{equation}\label{3.66}
\text{either }\ w_i(x)\equiv0\ \, \text{in}\ \, \R^3\ \ \    \text{or}\ \ \  	(\hat{\mu}_i, w_i)\ \text{ is\  an\  eigenpair\ of} \ \hat H_{\gamma}.
\end{equation}
Moreover, since $(\mu_i^{a_n},u_i^{a_n})$ is the $i$th eigenpair  (counted with multiplicity) of $H_V^{a_n}$ in $\R^3$, we deduce from \cite[Section 11.8]{analysis}
that $|u_1^{a_n}|>0$ holds in $\R^3$, and both $u_2^{a_n}$ and $u_3^{a_n}$ change sign.
Together with the fact that
$$
w_i(x)=\lim\limits_{n\to\infty}\epsilon_{a_n}^{\frac{3}{2}}u_i^{a_n}(\epsilon_{a_n} x+y_{k_*})\ \ \text{a.e. in} \ \, \R^3,\ \ i=1,2,3,
$$
one can further deduce from \eqref{3.66} that $w_1$ satisfies
\begin{equation}\label{3.75}
\text{either }\ w_1(x)\equiv0\ \, \text{in}\ \, \R^3\ \ \,   \text{or}\ \ \, 	(\hat{\mu}_1,w_1)\ \text{is\ the\ first\ eigenpair\ of}\ \hat H_{\gamma},
\end{equation}
and
\begin{equation}\label{3.67}
\text{both}\, \	(\hat{\mu}_2,w_2) \ \, \text{and}\, \ (\hat{\mu}_3,w_3)\ \text{cannot\ be\ the\ first\ eigenpairs\ of}\  \hat H_{\gamma},
\end{equation}
where the first eigenfunction is simple.

Since
\begin{equation}\label{3.74}
2=\text{Rank}(\!\gamma) =\text{Rank}\big(\sum_{i=1}^3|w_i\rangle\langle w_i|\big)=\text{dim}\big(\text{span}\big\{w_1, w_2,w_3\big\}\big),
\end{equation}
we next have the following two different cases:

{\em Case 1: $w_i(x)\not\equiv0$ in $\R^3$ holds for all $i=1,2,3$}.  In this case, since $w_i\not\equiv0$  in $\R^3$ holds for all $i=1,2,3$,  we conclude from \eqref{3.66}--\eqref{3.67} that
\begin{equation}\label{3.75a}
(\hat{\mu}_1,w_1)\ \text{is\ the\ first\ eigenpair\ of}\ \hat H_{\gamma},
\end{equation}
and
\begin{equation}\label{3.67a}
(\hat{\mu}_2,w_2) \ \, \text{and}\, \ (\hat{\mu}_3,w_3)\ \text{are\ the\ eigenpairs\ of}\  \hat H_{\gamma},\ \text{but\ not\ the\ first\ ones,}
\end{equation}
where $\hat H_{\gamma}$ is as in (\ref{3.55}).
Applying \cite[Section 11.8]{analysis} again, we derive from \eqref{3.75a} and \eqref{3.67a} that
\begin{equation}\label{3.59}
(w_1,w_2)=(w_1,w_3)=0.
\end{equation}
We thus  get from \eqref{3.74} and \eqref{3.59}  that  $w_3(x)\equiv tw_2(x)\not\equiv0$ in $\R^3$ for some $ t\in\R\backslash\{0\}$, which further yields that $\gamma$ can be simplified as
\begin{equation}\label{3.71}
\gamma=|w_1\rangle\langle w_1|+(1+t^2)|w_2\rangle\langle w_2|.
\end{equation}
Applying \eqref{3.59} and \eqref{3.71}, we then conclude from  \eqref{3.62} that
\begin{equation}\label{3.77b}
1=\|\gamma\|=\max\Big\{\|w_1\|_2^2,\  (1+t^2)\|w_2\|_2^2\Big\}.
\end{equation}
On the other hand,  by the fact that Rank$(\!\gamma) =2$, we further deduce from \eqref{form}, \eqref{3.62} and \eqref{3.71} that
\begin{equation}\label{3.77a}
2=\|\gamma\|\text{Rank}(\gamma) =\int_{\R^3}\rho_\gamma dx=\int_{\R^3}\big[w_1^2+(1+t^2)w_2^2\big]dx.
\end{equation}
We thus derive from \eqref{3.77b} and \eqref{3.77a} that
\begin{equation}\label{3.77}
1=\|\gamma\|=\|w_1\|_2^2= (1+t^2)\|w_2\|_2^2.
\end{equation}
Together with the fact that $w_3(x)\equiv tw_2(x)$ in $\R^3$, we conclude from \eqref{3.77} that
\begin{equation}\label{3.78}
w_3(x)\equiv \pm\sqrt{\|w_2\|_2^{-2}-1} \, w_2(x)\ \ \text{in} \ \, \R^3.
\end{equation}
Case 1 of Theorem \ref{th4} (2) is therefore proved in view of (\ref{3.59}), \eqref{3.77} and \eqref{3.78}.

{\em Case 2: There exists exactly one $i_*\in\{1,2,3\}$ such that $w_{i_*}(x)\equiv0$ in $\R^3$}.  In this case,  we derive from \eqref{form}, \eqref{3.42}, \eqref{3.62} and (\ref{3.74}) that
$$
0<\|w_i\|_2^2\leq1 \ \ \text{for} \ \ i\neq i_*, \ \   \mathrm{and}\ \   \sum_{i\neq i_*}^3\int_{\R^3}w_i^2dx=\int_{\R^3}\rho_\gamma dx=\|\gamma\|\, \mathrm{Rank}(\!\gamma) =2,
$$
which yield that  $\|w_i\|_2^2=1$ holds for all $i\neq i_*$. Together with \eqref{3.44}, this gives that
\begin{equation*}
w_i^{a_n}\to w_i \ \ \mathrm{  strongly \ in }\ H^1(\R^3)\ \ \mathrm{as}\ \ n\to\infty,\ \  i\neq i_*,
\end{equation*}
and hence
\begin{equation}\label{3.79}
(w_i,w_j)=\delta_{ij}\ \ \text{ for} \ \ i,j\in\{1,2,3\}\backslash\{i_*\}.
\end{equation}

We now claim that $i_*\neq1$. On the contrary, suppose $i_*=1$. Then $\gamma=\sum_{i=2}^3|w_i\rangle\langle w_i|$ is a minimizer
of $a_3^*$, which  satisfies $(w_i,w_j)=\delta_{ij}$ for $i,j=2,3$. This indicates from \cite[Theorem 6]{ii} that $(\mu_2, w_2)$ is the first eigenpair of the operator $\hat H_\gamma=-\Delta -\frac{5}{3}a^*_3\Big(\sum_{j=2}^3w_j^2
\Big)^{2/3}$ in $\R^3$, which however contradicts with  the fact \eqref{3.67}. Thus, the claim $i_*\neq1$ holds true. 

We finally prove that $i_*\neq2$. By contradiction, suppose $i_*=2$. We then have $w_2(x)\equiv0$ in $\R^3$. This implies from \eqref{3.74} that $\gamma=|w_1\rangle\langle w_1|+|w_3\rangle\langle w_3|$ is a minimizer of $a_3^*$ and  satisfies $(w_i,w_j)=\delta_{ij}$ for $i,j=1,3$. We thus obtain from \cite[Theorem 6]{ii}  that $\hat{\mu}_1<\hat{\mu}_3<0$, and $\hat{\mu}_1,\,\hat{\mu}_3 $ are the first two  eigenvalues 
of the operator $\hat H_\gamma=-\Delta -\frac{5}{3}a^*_3\Big(\sum_{j\neq2}^3w_j^2\Big)^{2/3}$ in $\R^3$.  Together with  \eqref{3.73}, we further derive that $\hat{\mu}_2<0$, and hence
\begin{equation}\label{3.87}
\sum_{i=1}^3\hat{\mu}_i<\sum_{i\neq2}^3\hat{\mu}_i.
\end{equation}

On the other hand, we calculate from \eqref{3.63a}, \eqref{3.72} and \eqref{3.73} that
\begin{equation}\label{3.88}
\begin{split}
\sum_{i=1}^3\hat{\mu}_i
&=\lim\limits_{n\to\infty}\sum_{i=1}^3\epsilon_{a_n}^2\mu^{a_n}_i\\
&=\lim\limits_{n\to\infty}\Big[\sum_{i=1}^3\int_{\R^3}|\nabla w_i^{a_n}|^2dx-\frac{5}{3}a_n\int_{\R^3}\Big(\sum_{i=1}^3|w_i^{a_n}|^2\Big)^{\frac{5}{3}}dx\Big],
\end{split}
\end{equation}
and
\begin{align}\label{3.89}
&\sum_{i\neq2}^3\hat{\mu}_i
=\lim\limits_{n\to\infty}\sum_{i\neq2}^3\epsilon_{a_n}^2\mu^{a_n}_i\nonumber\\
=&\lim\limits_{n\to\infty}\Big[\sum_{i\neq2}^3\int_{\R^3}|\nabla w_i^{a_n}|^2dx-\frac{5}{3}a_n\int_{\R^3}\Big(\sum_{i=1}^3|w_i^{a_n}|^2\Big)^{\frac{2}{3}}\Big(|w_1^{a_n}|^2+|w_3^{a_n}|^2\Big)dx\Big]\nonumber\\
=&\lim\limits_{n\to\infty}\Big[\sum_{i\neq2}^3\int_{\R^3}|\nabla w_i^{a_n}|^2dx-\frac{5}{3}a_n\int_{\R^3}\Big(\sum_{i=1}^3|w_i^{a_n}|^2\Big)^{\frac{5}{3}}dx\Big.\\
&\ \ \ \ \ \ \ \ \ \ \Big.+\frac{5}{3}a_n\int_{\R^3}\Big(\sum_{i=1}^3|w_i^{a_n}|^2\Big)^{\frac{2}{3}}|w_2^{a_n}|^2dx\Big].\nonumber
\end{align}
Applying \eqref{3.45}, \eqref{3.44} and the fact that $w_2(x)\equiv0$ in $\R^3$, one can further derive from \eqref{3.88} and \eqref{3.89} that
\begin{equation}\label{3.90}
\begin{split}
\sum_{i=1}^3\hat{\mu}_i
=\sum_{i\neq2}^3\int_{\R^3}|\nabla w_i|^2dx-\frac{5}{3}a^*_3\int_{\R^3}\Big(\sum_{i\neq2}^3w_i^2\Big)^{\frac{5}{3}}dx,
\end{split}
\end{equation}
and
\begin{align}\label{3.91}
\sum_{i\neq2}^3\hat{\mu}_i
=&\sum_{i\neq2}^3\int_{\R^3}|\nabla w_i|^2dx-
\frac{5}{3}a^*_3\int_{\R^3}\Big(\sum_{i\neq2}^3w_i^2\Big)^{\frac{5}{3}}dx\nonumber\\
&+\frac{5}{3}\lim\limits_{n\to\infty}a_n\int_{\R^3}\Big(\sum_{i=1}^3|w_i^{a_n}|^2\Big)^{\frac{2}{3}}|w_2^{a_n}|^2dx\\
:=&\sum_{i\neq2}^3\int_{\R^3}|\nabla w_i|^2dx-
\frac{5}{3}a^*_3\int_{\R^3}\Big(\sum_{i\neq2}^3w_i^2\Big)^{\frac{5}{3}}dx+\frac{5}{3}A.\nonumber
\end{align}
Note from \eqref{3.64} that
\begin{eqnarray*}
	w_2^{a_n}\to  w_2=0\ \ \  \ \mathrm{  strongly \ in}\ \ L^\infty(\R^3)\ \ \mathrm{as}\ \ n\to\infty,
\end{eqnarray*}
which yields that
\begin{equation}\label{3.92}
\begin{split}
0\leq A
:=&\lim\limits_{n\to\infty}a_n\int_{\R^3}\rho_{\tilde{\gamma}_{a_n}}^{ \frac{2}{3}}|w_2^{a_n}|^2dx\\
\leq&a_3^*\lim\limits_{n\to\infty}\|w_2^{a_n}\|_\infty\int_{\R^3}\rho_{\tilde{\gamma}_{a_n}}^{\frac{2}{3}}|w_2^{a_n}|dx\\
\leq&a_3^*\lim\limits_{n\to\infty}\|w_2^{a_n}\|_\infty\ \|\rho_{\tilde{\gamma}_{a_n}}\|_{4/3}^{2/3}\ \, \|w_2^{a_n}\|_2\\
=&0,
\end{split}
\end{equation}
i.e., $A=0$, where the last identity follows from the uniform boundedness of the sequence $\{\rho_{\tilde{\gamma}_{a_n}}\}$ $ =\big\{\sum_{i=1}^3|w_i^{a_n}|^2\big\}$ in $L^1(\R^3)\cap L^3(\R^3)$.
We thus conclude from \eqref{3.90}--\eqref{3.92} that
$$
\sum_{i=1}^3\hat{\mu}_i=\sum_{i\neq2}^3\hat{\mu}_i,
$$
which however contradicts with \eqref{3.87}. This proves that $i_* \not = 2$, and hence it necessarily has $i_* = 3$.  Case 2 of Theorem 1.3 (2) is therefore proved, and we are done.  \qed

\vspace{.5cm}

\noindent {\bf Data Availability:} Data sharing not applicable to this article as no datasets were generated or analysed during the current study.

\end{document}